\documentclass[12pt]{extarticle}
\pdfoutput=1 
\usepackage{slashed}
\usepackage{xcolor}
\usepackage{latexsym}
\usepackage{amssymb}
\usepackage{amsmath}
\usepackage{graphicx}
\usepackage{subcaption}
\usepackage{arydshln}
\usepackage{adjustbox}
\usepackage{easybmat}
\usepackage{bbm}
\usepackage{cite}
\usepackage{slashed}
\usepackage{bm}
\usepackage{adjustbox}
\usepackage{booktabs}
\usepackage{multirow} 
\usepackage{rotating}
\usepackage{mathtools}
\usepackage{lscape}
\usepackage{booktabs}
\usepackage{hyperref}
\usepackage{bm}
\usepackage{calrsfs}

\usepackage{bbold}
\usepackage[normalem]{ulem}
\newcommand{\trace}[1]{\left\langle #1 \right\rangle}

\newcommand{\rr}[0]{\tilde{r}}

\newcommand{\sumintB}[1]{{\hbox{$\sum$}\!\!\!\!\!\!\!\int\,}_{\!\!\!\!\!\!\raise-0.2ex\hbox{$\scriptstyle{#1}$}}}
\newcommand{\sumintF}[1]{{\hbox{$\sum$}\!\!\!\!\!\!\!\int\,}_{\!\!\!\!\!\!\raise-0.2ex\hbox{$\scriptstyle{\{#1\}}$}}}

\begin{document}
\thispagestyle{empty}
\begin{flushright}
\end{flushright}
\vspace{0.8cm}

\begin{center}
{\Large\sc Thermoskyrmions}
\vspace{0.8cm}

\textbf{
Mikael Chala~\footnote{mikael.chala@ugr.es}, Juan Carlos Criado~\footnote{jccriadoalamo@ugr.es
} and Luis Gil~\footnote{lgil@ugr.es}
}\\
\vspace{1.cm}
{\em {Departamento de F\'isica Te\'orica y del Cosmos, Universidad de Granada, Campus de Fuentenueva, E--18071 Granada, Spain}}\\[0.3cm]
\vspace{0.5cm}
\end{center}
\begin{abstract}
Skyrmions are stable and topologically non-trivial field configurations that behave like localized particles.
They appear in the chiral effective theory for pions, where they correspond to the baryon states, and might also exist in the electroweak theory, in the presence of certain effective interactions.
In this paper, focusing on toy models that capture different limits of the electroweak sector of the Standard Model (SM), we show that skyrmions not classically stable at zero temperature can be stabilized by thermal effects. This result motivates the study of skyrmions in the quantum effective action of the SM, potentially implying the existence of dark matter without new physics.

\end{abstract}

\setcounter{footnote}{0}

\newpage


\section{Introduction}

The interest in topological solitons in quantum field theory (QFT) dates back to the pioneering work of Skyrme \cite{Skyrme:1961vq}, who connected baryons to extended field configurations in chiral perturbation theory ($\chi$PT). These static configurations, later named \textit{skyrmions}, carry a conserved topological charge associated to their homotopy class, and are energetically stabilized by higher-derivative operators in the effective expansion. In \cite{Adkins:1983ya}, the relation between skyrmions and baryons was firmly established by Adkins, Nappi and Witten, who first quantized these solutions as fermions. Since these works, multiple theoretical connections between skyrmions and other non-trivial field configurations, such as Yang-Mills instantons, calorons and monopoles have been discovered (e.g. \cite{Atiyah:1989dq,Sutcliffe:2010et,Eskola:1989qk,Nowak:1989gw,Ward:2003sx,Cork:2018sem}), manifesting their ubiquity in QFT.

Beyond these intriguing aspects, other studies have brought attention to skyrmions in a cosmological context \cite{Murayama:2009nj,Joseph:2009bq,Gillioz:2010mr}. Similar to monopoles, skyrmions could be produced via the Kibble-Zurek mechanism \cite{Kibble:1976sj,Zurek:1985qw} during a cosmological phase transition with breaking pattern $G \to H$ with third homotopy group $\pi_3(G/H) = S^3$, or via skyrmion-antiskyrmion pair creation in very energetic collisions \cite{Bezrukov:2003er,Khoze:2020paj}. Their long typical lifetimes \cite{DHoker:1983ujb} make them appealing dark matter candidates. In this scenario, one requires beyond-the-SM (BSM) physics  that provides not only the required symmetry-breaking pattern above the electroweak (EW) scale, but also certain effective operators which make skyrmions energetically stable by virtue of Derrick's theorem \cite{Derrick:1964ww}. There are two main reasons why skyrmion configurations are hard to study in practice, assuming the theory presents the required non-trivial topological structure:

\begin{enumerate}
    \item \textbf{Classicality:} Skyrmions are classical solutions to the equations of motion with appropriate boundary conditions, and they must therefore be bosonic-field configurations. If the QFT in question contains fundamental fermions, one must find a regime of approximation where these degrees of freedom can be neglected or integrated out. In $\chi$PT this is natural, as the relevant degrees of freedom are mesons, and leptons do not couple to them at tree-level.
    \item \textbf{EFT validity:} The presence of higher-derivative operators competing against lower-derivative terms in the energy functional raises some skepticism about the validity of effective field theory (EFT) expansions in the study of skyrmions. If they are of similar size, the usual power counting, which typically holds if the UV completion is weakly coupled (like those described by the SMEFT \cite{Buchmuller:1985jz,Grzadkowski:2010es}), dictates that the contribution of operators with an arbitrarily large number of derivatives is unsuppressed\footnote{See \cite{Ellis:2012cs,Kitano:2016ooc,Kitano:2017zqw,Criado:2020zwu,Hamada:2021oqm,Criado:2021tec} for some recent studies on the existence and properties of skyrmions in the EW sector of the SMEFT and HEFT.}.
    This issue is circumvented if the EFT results from confinement at higher energies, as happens in $\chi$PT or composite Higgs models~\cite{Pomarol:2007kr,Domenech:2010aq,Gillioz:2011dj,Gillioz:2011tr,He:2017foi,Ellis:2012cs}, since the EFT expansion is not constructed by integrating out heavy degrees of freedom perturbatively.
\end{enumerate}

As we show in the present work, these two difficulties can be elegantly tackled in a high-temperature regime, such as the very early universe. The study of equilibrium phenomena in QFT at finite temperature can be carried out in the imaginary time formalism \cite{Matsubara:1955ws}, where Euclidean time is compactified to a circle of length $2 \pi / \beta$, where $\beta = 1/T$ is the inverse temperature. In this setup, bosons (fermions) satisfy periodic (antiperiodic) boundary conditions in the time direction, and can thus be decomposed in a tower of spatial Matsubara modes with heavy masses proportional to the temperature. In a high-temperature regime, all heavy Matsubara modes can be integrated out in favor of a static theory for the light zero Matsubara modes which live in three spatial dimensions (3D), and whose interactions encode all temperature dependence \cite{Ginsparg:1980ef,Appelquist:1981vg,Braaten:1995cm}. This procedure, called \textit{dimensional reduction}, is nowadays standard in the study of cosmological phase transitions~\cite{Chapman:1994vk,Brauner:2016fla,Andersen:2017ika,Niemi:2018asa,Gorda:2018hvi,Kainulainen:2019kyp,Gould:2019qek,Niemi:2020hto,Gould:2021ccf,Gould:2021dzl,Schicho:2021gca,Lofgren:2021ogg,Niemi:2021qvp,Camargo-Molina:2021zgz,Niemi:2022bjg,Ekstedt:2022bff,Ekstedt:2022ceo,Ekstedt:2022zro,Biondini:2022ggt,Gould:2022ran,Schicho:2022wty,Gould:2023jbz,Kierkla:2023von,Chala:2024xll,Niemi:2024axp,Qin:2024idc,Niemi:2024vzw,Camargo-Molina:2024sde,Gould:2024jjt,Chakrabortty:2024wto,Kierkla:2025qyz,Chala:2025oul,Bernardo:2025vkz,Chala:2025aiz,Chala:2025xlk,Keus:2025ova,Jahedi:2025yjz,Liu:2025ipj,Li:2025kyo,Annala:2025aci,Bhatnagar:2025jhh,Biekotter:2025npc,Chala:2025cya,Liu:2026ask} and hot QCD~\cite{Braaten:1995cm,Braaten:1994na,Braaten:1995jr,Kajantie:1997tt,Laine:2019uua,Laine:2018lgj,Ghiglieri:2021bom,Navarrete:2024ruu,Gorda:2025cwu}, as it enables perturbative calculations and bypasses numerous difficulties encountered within finite-temperature calculations in four spacetime dimensions (4D) \cite{Croon:2020cgk}. 

The resulting 3D EFT is purely bosonic, as antiperiodicity implies that all fermionic modes are heavy, then providing a clean setup to study skyrmion configurations of the bosonic zero modes. Furthermore, since the 3D EFT can be in principle constructed up to very high orders perturbatively, it is straightforward to assess the validity of the EFT expansion by ensuring the convergence of its contributions to the skyrmion properties. In the cases we are interested in here, the necessary stabilizing operators are generated in a thermal expansion, and we shall call the ensuing stable configurations \textit{thermoskyrmions}\footnote{We will use this name in order to distinguish them from the already studied thermal corrections to skyrmions in a heat-bath \cite{Dey:1994by}, where the stabilizing Skyrme operator is already present at $T=0$. In the models considered here, although we do not prove that stable topological configurations do not exist at $T=0$, they at least do not exist as classical configurations.}.

This article is organized as follows. In section \ref{sec:setup} we present a toy model that realizes the ideas above. In section \ref{sec:highT}, we describe the construction of its 3D EFT in the high-temperature limit. In section \ref{sec:skyrmions} we analyze the thermoskyrmion solutions of this theory as our main result, together with its thermodynamics. We conclude and point out possible future directions in section \ref{sec:conclusions}. We relegate some technical details to appendices \ref{app:matching} and \ref{app:operators}.

\section{Theoretical setup}
\label{sec:setup}

\subsection{Models}

We 
study a non-linear sigma model with global symmetry $G = SU(2)_L \times SU(2)_R$ and different interactions with one or two fermion doublets $\psi_{L(R)}$ under $SU(2)_{L(R)}$ and singlets of $SU(2)_{R(L)}$. We define the bi-doublet matrix
\begin{equation}
    U = \exp{\left(i \frac{\pi_a \sigma_a}{v}\right)}\,,
\end{equation}
where $a=1,2,3$ are $SU(2)$ indices, $\pi_a$ are real scalars, $\sigma_a$ stand for the Pauli matrices and $v$ is a normalization (decay) constant. %

In this notation, the relevant Lagrangian reads:
\begin{equation}\label{eq:UVlag}
    \mathcal{L} = \frac{v^2}{4} \trace{\partial_\mu U^\dagger \partial^\mu U} + \mathcal{L}_\mathrm{kin}^\psi - \mathcal{L}_\text{int}\,,
\end{equation}
where $\trace{\dots}$ amounts to tracing over $SU(2)$ indices, $\mathcal{L}_\mathrm{kin}^\psi$ accounts for the kinetic term of the fermions and $\mathcal{L}_\text{int}$, for any of the interactions including the fermions, as shown in Tab.~\ref{tab:models}.
\begin{table}
    \begin{center}
    \begin{tabular}{c|c|c}
        & Fermions & $\mathcal{L}_\text{int}$ \\[0.1cm]
        \hline
        &&\\[-0.3cm]
        Model A & $\psi_L = P_L \begin{pmatrix} \chi_1 \\ \chi_2 \end{pmatrix}$, $\psi_R = P_R \begin{pmatrix} \chi_1 \\ \chi_2 \end{pmatrix}$ & $\dfrac{y v}{\sqrt{2}} \overline{\psi}_L U \psi_R + \text{h.c.}$ \\[0.3cm]
        Model B & $\psi_L = P_L \begin{pmatrix} \chi_1 \\ \chi_2 \end{pmatrix}$ & $i g \overline{\psi}_L\gamma^\mu U \partial_\mu U^\dagger \psi_L$ \\[0.1cm]
        \hline
    \end{tabular}
    \end{center}
    \caption{\it Fermion content and interaction Lagrangian for the models discussed. $P_{L/R}$ stand for the left/right chirality projectors and $y, g$ are real-valued couplings.}\label{tab:models}
\end{table}

Both models can be understood as simplifications of a theory resembling the electroweak sector of the SM. Indeed, consider a $SU(2)_L$ gauge theory with global symmetry $SU(2)_L \times SU(2)_R$, whose field content consists of: gauge bosons $W_\mu$, a Higgs-like complex scalar bi-doublet matrix $H$ and fermion doublets $\psi_L$ and $\psi_R$. The most general Lagrangian reads:
\begin{align}
    \mathcal{L} &= -\frac{1}{4} \trace{W_{\mu \nu} W^{\mu \nu}} + \trace{(D_{\mu} H)^\dagger D^\mu H} + i \overline{\psi}_L \slashed{D} \psi_L + i \overline{\psi}_R \slashed{\partial} \psi_R \nonumber\\
    & - \left( y \overline{\psi}_L H \psi_R + \text{h.c.} \right) - V(H)\,,
    \label{eq:EW}
\end{align}
where $D_\mu = \partial_\mu - i g W_\mu^a \sigma_a$ is the $SU(2)_L$ gauge covariant derivative and $V(H)$ is the scalar potential.

After spontaneous symmetry breaking, the scalar acquires a vacuum expectation value (vev) $\sqrt{(H^\dagger H)} = v/\sqrt{2}$, so one can write:
\begin{equation}
    H(x) = \frac{h(x)+v}{\sqrt{2}} U(x)\,,
\end{equation}
where $h(x)$ is a real scalar singlet and $U(x)$ is a unitary matrix containing the Goldstone bosons, which transforms as a bi-doublet under $SU(2)_L \times SU(2)_R$. If the Higgs mass is very large, 
the radial component is effectively frozen to its vev ($h=0$), and the broken Lagrangian reads:
\begin{align}
    \mathcal{L}_\text{bro} &= -\frac{1}{4} \trace{W_{\mu \nu} W^{\mu \nu}} + \frac{v^2}{4} \trace{(D_{\mu} U)^\dagger D^\mu U} + i \overline{\psi}_L \slashed{D} \psi_L + i \overline{\psi}_R \slashed{\partial} \psi_R \nonumber\\
    & - \frac{g^2 v^2}{4} \trace{W_\mu W^\mu} - \left( \frac{y v}{\sqrt{2}} \overline{\psi}_L U \psi_R + \text{h.c.} \right)\,.
\end{align}
From here, one recovers model A by taking the limit where the $SU(2)_L$ gauge coupling vanishes, which is equivalent to setting $W_\mu = 0$. On the other hand, if  the Yukawa interaction is neglected ($y \to 0$),
the Lagrangian reads, in the unitary gauge\footnote{This model with a single left-handed doublet is known to suffer from a gauge $SU(2)$ anomaly \cite{Witten:1982fp} which makes it mathematically inconsistent. We shall not work with this model directly, as it was presented simply to give some physical motivation for our toy model with global $SU(2)$ symmetry.}:
\begin{equation}
     \mathcal{L}_\text{bro} = -\frac{1}{4} \trace{W_{\mu \nu} W^{\mu \nu}} + i \overline{\psi}_L \slashed{D} \psi_L - \frac{g^2 v^2}{4} \trace{W_\mu W^\mu}\,.
     \label{eq:broken}
\end{equation}
where $\psi_R$ is not included, as it is non-interacting. Now, we recover model B if $W_\mu$ takes on a pure gauge configuration, $W_\mu = \frac{i}{g} (\partial_\mu U) U^\dagger$, in which case the field strength vanishes.
It is reasonable to assume that, if a skyrmion solution is present in the theory, it must be close to a pure gauge configuration, since the term $\langle W_{ij} W^{ij}\rangle$ (with $i = 1, 2, 3$) tends to increase the energy away from a pure gauge.
This is typically the case in models of the electroweak sector~\cite{Ambjorn:1984bb, Criado:2020zwu, Hamada:2021oqm, Criado:2021tec}.

\subsection{Skyrmion configurations}

For static configurations of the fields $f$ ($\partial_0 f = 0$), the energy density is given by the Hamiltonian density $\mathcal{H} = - \mathcal{L}$, where the Lagrangian $\mathcal{L}$ is a function of static fields only. The energy functional for a given static configuration of $U$, $U_0(\mathbf{x}) \equiv U(t=t_0, \mathbf{x})$ is
\begin{equation}
    E[U_0] = \int d^3 x \mathcal{H}(U_0)\,.
\end{equation}
Our goal is to find the $U_0(\mathbf{x})$ that minimizes the energy functional in models A and B. Such configuration should have finite energy, which requires that $U(\mathbf{x})$ be continuous and that its spatial derivatives vanish at infinity. Therefore, $SU(2)$ symmetry can be used to choose this constant element at infinity to be the identity, i.e. $\lim_{|\mathbf{x}| \to \infty} U_0(\mathbf{x}) = \mathbb{1}_2$. Therefore, $U_0$ defines a continuous mapping
\begin{equation}
    U_0: \mathbb{R}^3 \cup \{\infty\} \cong S^3 \to SU(2) \cong S^3\,,
\end{equation}
where $\cong$ denotes an homeomorphism relation. Such mappings can be classified according to an integer winding number, since the third homotopy group of the 3-sphere is $\pi_3(S^3) = \mathbb{Z}$. This winding number is defined as
\begin{equation}\label{eq:nU}
    n_U = -\frac{1}{24 \pi^2} \epsilon_{ijk} \int d^3 x \trace{(U^\dagger \partial_i U) (U^\dagger \partial_j U) (U^\dagger \partial_k U)}\,,
\end{equation}
where $\epsilon_{ijk}$ is the fully antisymmetric tensor with spatial indices. A winding number is an homotopy invariant, i.e. it is unchanged under continuous deformations of $U_0(\mathbf{x})$, and in particular under time evolution. A trivial vacuum configuration, $U_0(\mathbf{x}) = \mathbb{1}_2$ has $n_U = 0$, so any non-trivial configuration with $n_U > 0$ is said to be \textit{topologically protected}, in the sense that it cannot decay into the vacuum. 

The fact that $n_U$ is an homotopy invariant is not a necessary nor a sufficient condition for the existence of skyrmions. In order for an extended field configuration to be a (stable) minimum of the energy functional, Derrick's theorem \cite{Derrick:1964ww} states that the Hamiltonian must contain operators with at least four derivatives that can stabilize its size. %

In models A and B, the question of stability is impossible to address by analyzing the theory at the classical level, as originally done in $\chi$PT, for instance. The reason is that the Hamiltonian is not only a function of $U$, but also of Grassmann-valued fermions $\psi_L$ and $\psi_R$, which cannot take on a classical configuration. If one insists on determining whether $U$ skyrmions exist in these theories, one would in principle need to construct a one-loop quantum effective action $\Gamma[U]$ for the bosonic field.
Furthermore, it is not very clear if any relevant physical conclusion can be drawn from such construction, or even if the problem is amenable to numerical computations. 

Only in the case that fermions have very large masses $m_\psi$, the quantum effective action can be realized as a local EFT expansion in inverse powers of $m_\psi$. Higher-derivative operators are naturally generated in the EFT for the light Goldstones $\pi_a$ in $U$, and the energy can be numerically minimized to look for skyrmions which are solutions for the effective, quantum corrected field $U$. This is however not the case in model A if the Yukawa $y$ is sufficiently small, since $m_\psi = y v / \sqrt{2}$, nor in model B, where fermions are taken to be massless.

As we show in the next section, this is not an issue if 
the theory is studied at finite temperature.

\subsection{Thermal field theory}
\label{sec:tft}

At finite temperature in equilibrium, instead of the generating functional of QFT, the goal is to compute the canonical partition function $\mathcal{Z}$, which serves an analogous purpose in the determination of relevant thermodynamical variables \cite{Laine:2016hma}. Indeed, if $H$ is the Hamiltonian and $q$ is a generalized coordinate, one can write \cite{Matsubara:1955ws}
\begin{equation}
    \mathcal{Z} = \sum_q \langle q;0| \exp\left(-\beta H \right) | q;0 \rangle  = \int_{q(0) = q(-i \beta)} \mathcal{D}q~\exp\left(-S_E\right)\,.
    \label{eq:generating functional th}
\end{equation}
Note that in this path integral, the action $S_E$ is Euclidean and $q(t)$ is periodic in Euclidean time, after a Wick rotation $\tau = i t$, with a period equal to $\beta=1/T$, where $T$ is the temperature. Basically, a QFT at finite temperature is equivalent to an Euclidean QFT at zero temperature where the time dimension is compactified to a circle of length $\beta = 1/T$.

The compactification in the time direction $\mathbb{R}^{3,1} \to \mathbb{R}^3 \times S^1$ leads to the splitting of fields into a tower of discrete thermal modes. For bosonic fields one has
\begin{equation}
    \phi(\tau, \mathbf{x}) \equiv \sum_{n=-\infty}^\infty \phi_n(\mathbf{x}) e^{i \omega_n \tau}\,,
    \label{eq:modes scalar}
\end{equation}
where periodicity imposes that $\omega_n = 2 \pi n T$. These frequencies are called \textit{Matsubara frequencies}, and the corresponding amplitudes $\phi_n(\mathbf{x})$ are called \textit{Matsubara modes}. Analogously, fermions satisfy
\begin{equation}
    \psi(\tau, \mathbf{x}) \equiv \sum_{n=-\infty}^\infty \psi_n(\mathbf{x}) e^{i \omega'_n \tau}\,,
    \label{eq:modes fermion}
\end{equation}
where antiperiodicity (due to their anticommutation) implies that $\omega'_n = 2 \pi (n+\frac{1}{2}) T$.

In the Matsubara formalism, the (Euclidean) momentum of each Matsubara mode is $P=(\omega_n, \mathbf{p})$ for bosons and $P=(\omega'_n, \mathbf{p})$ for fermions. For each mode, the on-shell relation for 3D momenta is $P^2 = \omega_n^2 + |\mathbf{p}|^2 = -m^2$ (resp. $\omega_n^{' 2}$). The sum $m_n^2 \equiv m^2 + \omega_n^2$ (resp. $\omega_n^{' 2}$) is known as the \textit{thermal mass}. As a result, loop integrals are replaced by sum-integrals, which are infinite sums over Matsubara modes including integration over 3D momenta:
\begin{equation}
    \int \dfrac{d^4 q}{(2 \pi)^4} \to \sumintB{Q} \equiv T \sum_{n=-\infty}^\infty \int \dfrac{d^3 q}{(2 \pi)^3} \,.
\end{equation}

In a high-temperature regime, the hierarchy of scales between the light bosonic zero modes ($n=0$) and heavy thermal modes ($|n|>0$ for bosons and all $n$ for fermions) motivates an EFT treatment \cite{Kajantie:1995dw}. Through dimensional reduction (DR), one can build a 3D spatial EFT for the bosonic zero modes only, where the effect of heavy thermal modes enters the EFT through a modification of the Wilson coefficients (WC) of zero modes, which capture all temperature dependence. In practice, this is done using the standard matching techniques for $T=0$ EFT, with the exception that loop integrals are replaced by sum-integrals and $SO(1,3)$ invariance is broken to $SO(3)$ in the spatial sections.

Going back to our original purpose, that is the computation of skyrmion configurations, we note that the high-temperature limit of a QFT is necessarily bosonic, as all thermal modes of fermions become heavy and decouple for large-enough temperature. If 
we perform DR in our models, we are left with a theory for the zero mode of the $\pi_a$ (which are encoded in a $SU(2)$ matrix that we shall call $U$ as well, with a little abuse of notation) that is semiclassical, in the sense that fermions $\psi_{L,R}$ and heavy modes of the $\pi_a$ are accounted for through their quantum corrections to the zero mode of the $\pi_a$, but the theory is kept classical\footnote{Further quantum corrections from bosonic zero modes may also be computed, and has in fact been done so to high orders in the context of cosmological phase transitions (see e.g. \cite{Ekstedt:2024etx}).} in the zero mode of the $\pi_a$. 

The partition function in the high-temperature limit, once the heavy $\pi T$ scale is integrated out to construct a 3D EFT for the zero mode of the $\pi_a$, reads
\begin{equation}
    \mathcal{Z} = \int^{(\pi T)} \mathcal{D} U ~ e^{-S_3[U](T)}\,,
\end{equation}
where the 3D effective action is
\begin{equation}
    S_3[U](T) = \int d^3 x ~\mathcal{L}_3(U, \partial_i U; T)\,,
\end{equation}
and all temperature dependence is encoded in the WCs in the 3D Lagrangian $\mathcal{L}_3$. In its effective expansion, this Lagrangian can contain the required local, stabilizing operators for skyrmions. In such case, the large temperature scale would be responsible for their existence as semiclassical configurations, as our models do not contain such static configurations at $T=0$. Therefore, if skyrmions exist in this situation, we shall call these thermally stabilized, classical configurations thermoskyrmions (TS). We shall return to the discussion of the role of the temperature in the existence of such configurations in section \ref{sec:T role}.

In what follows, we construct the high-temperature limits of models A and B and study whether TS solutions can arise in these theories.

\section{High-temperature limit}
\label{sec:highT}

Since the 3D EFTs for models A and B contain the same degrees of freedom $U$ (non-linear representation of the zero modes of the $\pi_a$) and the same symmetries (global group $SU(2)_L \times SU(2)_R$ and rotational invariance), we use the same basis of operators in both cases. At $\mathcal{O}(p^2)$, the most general basis contains a single operator:
\begin{equation}\label{eq:lagp2}
    \mathcal{L}_3^{(p^2)} = c_0 T \trace{\partial_i U^\dagger \partial_i U}\,,
\end{equation}
where we have factored out a power of $T$ to make $c_0$ dimensionless. At $\mathcal{O}(p^4)$, one has \cite{Scherer:1994wi}:
\begin{align}\label{eq:lagp4}
    \mathcal{L}_3^{(p^4)} &= \frac{c_1}{T} \trace{\partial_i U^\dagger \partial_i U}^2 + \frac{c_2}{T} \trace{\partial_i U^\dagger \partial_j U} \trace{\partial_i U^\dagger \partial_j U} + \textcolor{gray}{\frac{c_3}{T} \trace{(\partial^2 U U^\dagger - U \partial^2 U^\dagger)^2}}\,,
\end{align}
where the particular combination $c_2 = -c_1$ yields the well-known Skyrme operator \cite{Skyrme:1961vq}.
The term in gray can be eliminated through an appropriate field redefinition; we say that it is \textit{on-shell redundant}. As detailed in \cite{Scherer:1994wi}, at lowest order one can consider the equation of motion (EOM) from Eq.~\eqref{eq:lagp2}, $\partial^2 U U^\dagger - U \partial^2 U^\dagger = 0$, to see that this operator does not contribute to the action for physical configurations of $U$. 

As we show below, we also need the $\mathcal{O}(p^6)$ operators for model B. The basis is not as simple at $\mathcal{O}(p^6)$ (see \cite{Fearing:1994ga}), and it was shown in \cite{Ruiz-Femenia:2015mia} that there are six on-shell-independent operators. In principle, as found in \cite{Fearing:1994ga}, there is also a set of seven redundant operators that are required for matching and also vanish upon using field redefinitions. The following receive a non-vanishing matching contribution:
\begin{align}\label{eq:lagp6}
    \mathcal{L}_3^{(p^6)} &= \frac{B_{53}}{T^3} \mathcal{O}_{53} + \frac{B_{54}}{T^3} \mathcal{O}_{54} + \frac{B_{55}}{T^3} \mathcal{O}_{55} + \frac{B_{58}}{T^3} \mathcal{O}_{58} + \frac{B_{100}}{T^3} \mathcal{O}_{100} + \frac{B_{101}}{T^3} \mathcal{O}_{101} \nonumber\\
    & + \textcolor{gray}{  \frac{E_{2}}{T^3} \mathcal{O}_{2} + \frac{E_{3}}{T^3} \mathcal{O}_{3}}\,.
\end{align}
We use the same numerical tags for physical ($B_n$) and redundant ($E_n$) operators $\mathcal{O}_n$ as those used in Tables II - VIII in \cite{Fearing:1994ga}, and their expressions can be found in appendix \ref{app:operators}. 

Therefore, the 3D Lagrangian in Euclidean signature reads:
\begin{equation}
    \mathcal{L}_3 = \mathcal{L}_3^{(p^2)} + \mathcal{L}_3^{(p^4)} + \mathcal{L}_3^{(p^6)}\,,
\end{equation}
where all coefficients above can be determined through matching as functions of the parameters in models A and B ($v, T, g, y$).

Let us also note that matching a non-linear field theory onto its 3D EFT is not straightforward, as effective operators are functions of the $U$, but it is the Goldstone constituents $\pi_a$ that appear in external legs of one-particle-irreducible (1PI) diagrams. For brevity, we do not include a detailed description of the matching in the main text, and present the 1-loop matching results directly. We refer the interested reader to appendix \ref{app:matching}, where we provide a detailed explanation and example of our approach. 

\subsection{Model A}

In the matching, 1-loop contributions from heavy thermal modes of $U$ only modify the kinetic term, and they do not generate any finite contribution to the higher-derivative operators. Fermions, however, do generate higher-derivative operators at different orders in the Yukawa $y$.

The leading term in the linear expansion of $U$ in $\pi_a$ reveals that fermions in model A have a mass $m_\psi = y v/\sqrt{2}$. As a result, apart from the vertices, the hard-region expansion \cite{Beneke:1997zp} of loop sum-integrals yields additional powers of the Yukawa as
\begin{equation}\label{eq:hrexpansion}
    \sumintF{Q} \frac{1}{Q^2 + m_\psi^2} = \sumintF{Q} \left[ \frac{1}{Q^2} - \frac{m_\psi^2}{Q^4} + \dots \right]\,.
\end{equation}
Therefore, the matching at each order in $y$ 
also requires an expansion in masses up to the appropriate order.

Four-derivative operators in model A are first generated at 1-loop at $\mathcal{O}(y^4)$, and no operators at six-derivatives appear at this order. Using the notation introduced in Eqs. \eqref{eq:lagp2} and \eqref{eq:lagp4}, the matching equations for the 3D WCs at this order read:
\begin{align}
    c_0 &= \frac{v^2}{4 T^2} \left( 1 - \frac{T^2}{6 v^2} - \frac{28 \zeta(3) y^4}{(4 \pi)^4} \frac{v^2}{T^2} \right)
    \,, \label{eq:modelA c0}\\
    c_1 &= - \frac{7 \zeta(3) y^2}{6 (4 \pi)^4} \frac{v^2}{T^2} + \frac{155 \zeta(5) y^4}{4 (4 \pi)^6} \frac{v^4}{T^4} 
    \,, \label{eq:modelA c1}\\
    c_2 &= - \frac{31 \zeta(5) y^4}{2 (4 \pi)^6} \frac{v^4}{T^4} 
    \,, \\
    \textcolor{gray}{c_3} &= \textcolor{gray}{\frac{7 \zeta(3) y^2}{12 (4 \pi)^4} \frac{v^2}{T^2}- \frac{31 \zeta(5) y^4}{4 (4 \pi)^6} \frac{v^4}{T^4}}
    \,;
\end{align}
where the bosonic contribution to $c_0$ agrees with the $\mathcal{O}(T^2)$ thermal contribution to the pion decay constant in two-flavors $\chi$PT originally computed in \cite{Eletsky:1993hp}.

At first glance, the physical coefficients $c_1$ and $c_2$ seem to be both negative. The associated operators can be seen to give positive contributions to the 3D action\footnote{See for instance $Q_{D1}$ and $Q_{D2}$ in Table 1 in \cite{Criado:2021tec}, which enter with a relative minus sign in the Hamiltonian in that work.}, which is functionally equivalent to the energy at $T=0$. According to Derrick's criterion (see Eqs. \eqref{eq:derrick1} and \eqref{eq:derrick2} below), in this situation the four-derivative terms are not stabilizing, and thus no non-trivial minimum of the action exists.

\begin{figure*}[t]
    \centering
    \begin{subfigure}[t]{0.49\textwidth}
        \centering
        \includegraphics[width=\textwidth]{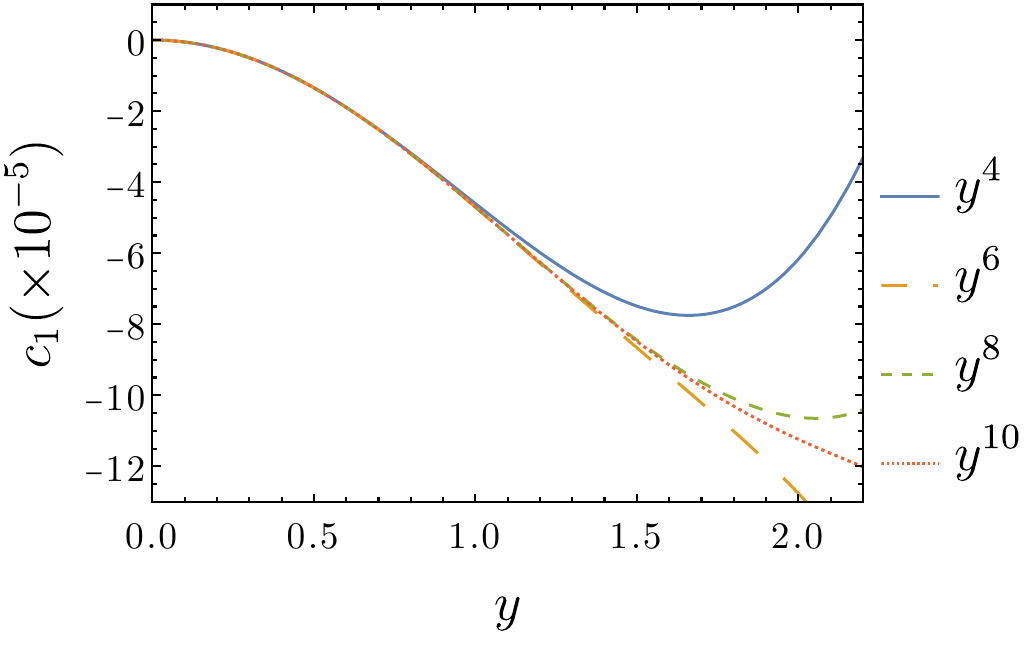}
    \end{subfigure}%
    ~ 
    \begin{subfigure}[t]{0.49\textwidth}
        \centering
        \includegraphics[width=\textwidth]{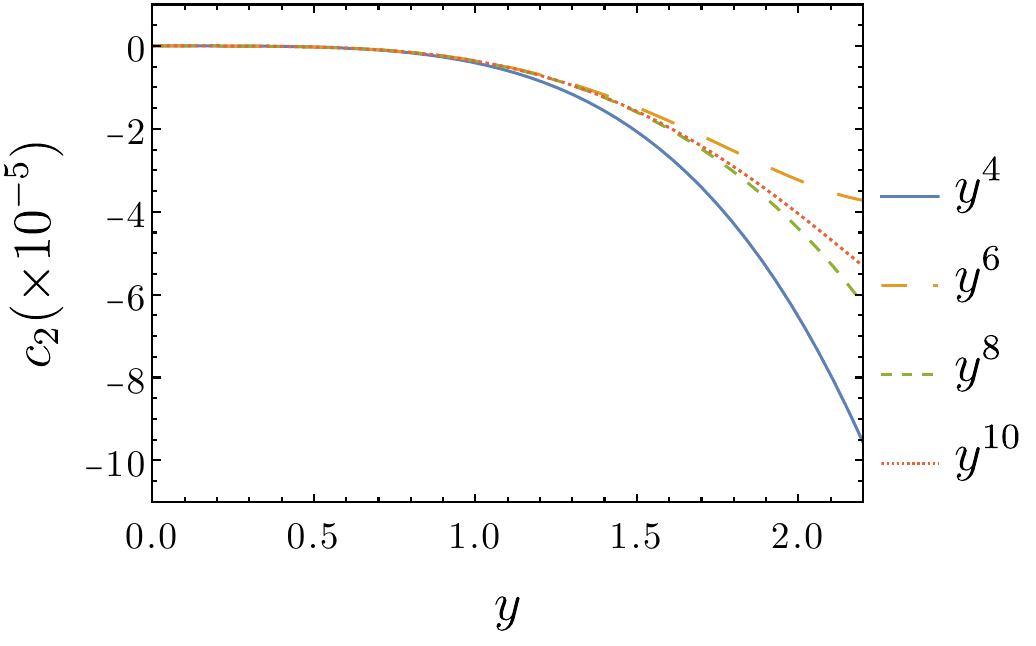}
    \end{subfigure}
    \caption{\it Four-derivative physical coefficients $c_1$ and $c_2$ in model A as a function of the Yukawa for fixed $T/v = 1$. The different curves show the 1-loop matching result at different orders in the Yukawa.
    }\label{fig:yukawas}
\end{figure*}

The second term in Eq. \eqref{eq:modelA c1} (and other higher powers in the Yukawa), however, might make the four-derivative contribution positive for large enough Yukawa. We inspect this in Figure~\ref{fig:yukawas}, where we show the convergence of the 1-loop matching equations up to $\mathcal{O}(y^{10})$ for $c_1$ and $c_2$. We see that the expansion converges as long as $y \lesssim 1.5$, and that within that regime both $c_1$ and $c_2$ remain negative throughout.

\subsection{Model B}

In this model with massless fermions, the matching at each loop order contains a finite power of the $g$ coupling. The kinetic term then only receives 1-loop Goldstone corrections and again reads~\footnote{That fermion matching corrections to the kinetic term are absent holds at any loop order. It can be understood from the fact that the fermion sector is invariant under a local $SU(2)$ transformation $U\to\Omega(\mathbf{x})U$ and $\psi_L\to \Omega(\mathbf{x})\psi_L$ with $\Omega$ unitary, whereas the scalar kinetic term is not.}:
\begin{equation}\label{eq:c0 model B}
    c_0 = \frac{v^2}{4 T^2} \left( 1 - \frac{T^2}{6 v^2} \right)\,.
\end{equation}
The interaction between $\psi$ and $U$ is also such that, at 1-loop, it simply renormalizes the four-derivative operators (they only yield a divergent $1/\epsilon$ term in dimensional regularization).  Upon inserting the appropriate counterterms, four-derivative WCs are then vanishing; so at 1-loop, the next order in the EFT expansion, namely six-derivative operators, must be explored. The latter do receive a finite contribution which, using the same notation as in Eq. \eqref{eq:lagp6}, reads:
\begin{align}
    B_{53} &= -\frac{7 \zeta(3) g^2 (1 + 2 g)^2}{80 \pi^4}\,,&
    B_{54} &= \frac{7 \zeta(3) g^2 (1 + 2 g)^2}{80 \pi^4}\,,
    \\[0.2cm]
    B_{55} &= -\frac{7 \zeta(3) g^2 (1 + 2 g)^2}{240 \pi^4}\,,&    
    B_{58} &= \frac{7 \zeta(3) g^2 (1 + 2 g)^2}{120 \pi^4}\,,
    \\[0.2cm]
    B_{100} &= \frac{7 \zeta(3) (1 + 6 g + 8 g^2)^2}{240 \pi^4}\,,&
    B_{101} &= -\frac{7 \zeta(3) (1 + 6 g + 8 g^2)^2}{240 \pi^4}\,,
    \\[0.2cm]
    \textcolor{gray}{E_{2}} &= \textcolor{gray}{-\frac{7 \zeta(3) g^2 (1 + 2 g)^2}{240 \pi^4}} \,,&
    \textcolor{gray}{E_{3}} &= \textcolor{gray}{-\frac{7 \zeta(3) g^2 (1 + 2 g)^2}{120 \pi^4}} \,.
\end{align}
We observe that these contributions present different signs, and given the complicated structure that arises from the sum of the corresponding operators (see \cite{Fearing:1994ga}) it is not clear a priori whether these can stabilize extended configurations from Derrick's criterion. We 
study the existence of stable configurations in this model in the next section.

\section{Thermoskyrmions}
\label{sec:skyrmions}

The simplest non-trivial TS configurations, distinct from the vacuum ($n_U=0$), are the single TS and anti-TS with unit winding numbers, $n_U=1$ and $n_U = -1$, respectively. These are found assuming the hedgehog ansatz \cite{Skyrme:1961vq}:
\begin{equation}\label{eq:hedgehog}
    U(\mathbf{x}) = \exp \left\{ i \eta(r) n_a \sigma_a \right\} \equiv U_\mathrm{TS}(\mathbf{x})\,,
\end{equation}
where $n_a$ are the components of the unit position vector, $\sigma_a$ are the Pauli matrices and $\eta(r)$ is the unknown radial profile. The name \textit{hedgehog} is due to the fact that in this configuration the Goldstone fields $\pi_a = \sin(\eta) n_a$ point radially outward from the origin at all points.

Within this ansatz, we impose that $U_\mathrm{TS}(\mathbf{x})$ is smooth at $\mathbf{x} = 0$ and its energy is finite.
The latter implies that $U_\mathrm{TS}(\mathbf{x})$ is a constant at spatial infinity.
Since a global rotation can be used to set this constant to any $SU(2)$ matrix, we choose $U_\mathrm{TS}(\mathbf{x}) = \mathbb{1}_2$ without loss of generality.
These two conditions translate into the following boundary conditions for the radial profile:
\begin{equation}
    \eta(0) = n \pi\, ~~(n \in \mathbb{Z})\,, \quad \eta(\infty) = 0\,.
\end{equation}
Upon inserting the ansatz into the expression for the winding number \eqref{eq:nU}, one gets:
\begin{equation}
    n_U = -\frac{2}{\pi} \int_0^\infty dr \partial_r \eta \sin^2(\eta) = \frac{1}{\pi} \eta(0) = n\,,
\end{equation}
so the winding number is determined by the boundary condition at the origin. In our case, we
then impose $\eta(0) = \pi$ for configurations with $n_U=1$.

In analogy with the electrical charge, $n_U$ allows us to define a natural size for a TS by averaging over the $n_U$ density:
\begin{equation}\label{eq:radius}
    R_\mathrm{TS}^2 = -\frac{2}{\pi} \int_0^\infty dr r^2 \partial_r \eta \sin^2(\eta)\,.
\end{equation}

In order to make the problem of finding an action-minimizing configuration more amenable to numerical methods, we perform a rescaling of the action functional. Formally, the 3D action of a radial profile $\eta(r)$, which depends on $\eta$ and its derivatives (generically denoted by $\partial_r \eta$) reads
\begin{equation}
    S_3[\eta] = \int d^3 x \mathcal{L}_3(\eta, \partial_r \eta) = 4 \pi \int dr r^2 \mathcal{L}_3(\eta, \partial_r \eta) \,.
\end{equation}
In terms of a dimensionless radius $\rr = v_3^2 e r$ ---where 
$v_3^2 \equiv 4 c_0 T$ is related to the 3D decay constant as shown in 
Eq. \eqref{eq:c0 model B} and has units of square root of energy, and $e$ is a dimensionless scaling parameter---, we have
\begin{equation}\label{eq:dimless action}
    S_3[\eta] = \frac{4 \pi}{(v_3^2 e)^3} \int d\rr \rr^2 \mathcal{L}_3(\eta, v_3^2 e \partial_{\rr} \eta) = \frac{4 \pi}{e} \int d\rr \rr^2 \frac{1}{v_3^6 e^2} \mathcal{L}_3(\eta, v_3^2 e \partial_{\rr} \eta) \equiv \frac{4 \pi}{e} \widetilde{S}_3[\eta] \,.
\end{equation}
%
$\widetilde{S}_3[\eta]$ is now an integral in a dimensionless variable and can be used to derive the TS solution numerically. We note that the $e$ parameter can be tuned to adjust the relative size of the integrands with different energy scalings (different number of derivatives), and the original 3D action can be recovered by performing the inverse rescaling at the end of the computation. 

Solving the EOM in the presence of higher-derivative operators while configurations are constrained to boundary conditions is however not an easy task for conventional methods. For this purpose, and in line with \cite{Criado:2020zwu, Criado:2021tec}, we opt for the direct minimization of the action using a neural network approach with \texttt{Elvet} \cite{Araz:2021hpx}. 

Our goal is to find a minimum of the 3D action for given values of $v, T, g$ and the parameter $e$. We model the profile $\eta(r)$ as a neural net with a single 10-unit layer, and the net is trained using the Adam minimization algorithm \cite{Kingma:2017} with loss functional
\begin{equation}\label{eq:loss}
    L[\eta] = \widetilde{S}_3[\eta] + \omega_\mathrm{BC} \sum_k \text{BC}_k[\eta]^2\,.
\end{equation}
Here, $\widetilde{S}_3[\eta]$ is the rescaled action in Eq. \eqref{eq:dimless action}, and $\text{BC}[\eta] \equiv \{\eta(0) - \pi, \eta(\rr_\mathrm{max}) = 0\}$ defines the boundary conditions with some hyperweight $\omega_\mathrm{BC}$ that is set by hand to ensure the conditions are satisfied. The integral for the action is computed by averaging over 1000 points uniformly distributed between $\rr=0$ and $\rr=5$, and the net is trained over a large number of epochs ($\sim 10^5$ in our study), until the loss functional converges.

To further ensure that the solution is a true minimum of the action, we check that Derrick's criterion \cite{Derrick:1964ww} is satisfied. As an illustrating example, assume that the action consists of a two-derivative term $\widetilde{S}_3^{(p^2)}$ and a four-derivative term $\widetilde{S}_3^{(p^4)}$. Upon dilation of the radial coordinate $\rr \to \lambda \rr$, the action becomes 
\begin{equation}
    \widetilde{S}_{3,\lambda} = \lambda \widetilde{S}_3^{(p^2)} + \frac{1}{\lambda} \widetilde{S}_3^{(p^4)}\,,
\end{equation}
and stability requires that
\begin{align}
    \frac{d \widetilde{S}_{3,\lambda}}{d \lambda} \biggr|_{\lambda = 1} &= \widetilde{S}_3^{(p^2)} - \widetilde{S}_3^{(p^4)} = 0\,, \label{eq:derrick1}\\
    \frac{d^2 \widetilde{S}_{3,\lambda}}{d \lambda^2} \biggr|_{\lambda = 1} &= 2 \widetilde{S}_3^{(p^4)} > 0\,. \label{eq:derrick2}
\end{align}
This shows why higher-derivative terms are needed in order for stable minima to exist, and also that these terms must have the correct relative sign to counterbalance the kinetic term. In practice, we ensure that the solutions we find are minima of the action by checking that the following ratio is at most at the percent level:
\begin{equation}
    \Delta = \frac{1}{\widetilde{S}_3}\frac{d \widetilde{S}_{3,\lambda}}{d \lambda} \bigg|_{\lambda = 1} \lesssim 10^{-2}\,.
\end{equation}

Now that we have presented our method to compute topological configurations that minimize the 3D action, let us inspect whether these solutions exist for model B.

\subsection{Finding a stable topological configuration}

In order to define a loss functional in terms of the radial profile $\eta(r)$, the 3D action in model B must first be expressed in the heghehog ansatz \eqref{eq:hedgehog}. The rescaled action $\widetilde{S}_3 = \widetilde{S}_3^{(p^2)} + \widetilde{S}_3^{(p^6)}$ in this ansatz is, at $\mathcal{O}(p^2)$:
\begin{align}
    \widetilde{S}_3^{(p^2)} = \int_0^\infty d\rr \left( \frac{1}{2} \rr^2 (\eta')^2 + \sin^2(\eta)  \right)\,,
\end{align}
where we now use $\eta' \equiv \partial_{\rr} \eta$ to lighten the notation. The $\mathcal{O}(p^6)$ piece is:
\begin{equation}\label{eq:B action p6}
    \widetilde{S}_3^{(p^6)} = \frac{7 \zeta(3) g^2 (1 + 2 g)^2}{15 (2\pi)^4} v_3^6 e^4 \int_0^\infty d\rr I^{(p^6)}(\rr)\,,
\end{equation}
where
\begin{align}
    I^{(p^6)}(\rr) &\equiv \frac{(\eta')^2 \sin^4(\eta)}{\rr^2} \biggl\{ 7 + 8 \rr^2 (\eta')^2 + 4 \rr^4 \biggl[(\eta'')^2 + (\eta')^4\biggr] 
    \nonumber \\
    &- 2 \biggl[5 + 4 \rr^2 (\eta')^2 \biggr] \cos(2 \eta) - 4 \rr \biggl( 5 \eta' + 2 \rr \eta'' \biggr) \sin(2 \eta) \nonumber \\
    & + \biggl[3 - 4 \rr^2 \biggl((\eta')^2 + 4 \rr \eta'' \eta' + \rr^2 \biggl( (\eta'')^2 - (\eta')^4 \biggr) \biggr)\biggr] \cos(4 \eta)  \nonumber \\
    & + 4 \rr \biggl[ 2 \rr \eta'' + \eta' \biggl( 3 + 2 \rr^2 \eta' \biggl( 2 \eta' + \rr \eta'' \biggr) \biggr) \biggr] \sin(4 \eta) \biggr\}\,.
\end{align}

Despite the apparent complexity of the integrand above, it is remarkable that all dependence on the coupling $g$ factorizes out of the integral. This is not necessary, but it turns out to be extremely convenient for our numerical approach, since an appropriate second rescaling of the radial coordinate allows us to write the action as an integral which is independent of all input parameters in the theory, namely $v, T$ and $g$. Explicitly, if we perform $\rr \to \rr' \equiv \lambda \rr$
\begin{equation}
    \widetilde{S}_3 \to \lambda \left( \widetilde{S}_3^{(p^2)} + \frac{1}{\lambda^4}\widetilde{S}_3^{(p^6)}\right) \equiv \lambda \widetilde{S}'_3\,,
\end{equation}
and set $\lambda^4 = \frac{7 \zeta(3) g^2 (1 + 2 g)^2}{15 (2\pi)^4} \frac{v_3^6}{T^3}$, then we cancel the prefactor in front of the integral in Eq. \eqref{eq:B action p6} (except for the $e^4$) and all dependence on the input parameters disappears. The new rescaled action $\widetilde{S}'_3$ can be minimized only once for an appropriate value of $e$ that makes the action take $\mathcal{O}(1)$ values, and the model can be studied for different input parameters by simply changing the prefactor in front the action. The same logic can be followed to derive a dimensionless radius from Eq. \eqref{eq:radius}, using the same rescaled variables of integration:
\begin{equation}
    R_\mathrm{TS}^2 \equiv \frac{1}{(v_3^2 e)^2} \widetilde{R}_\mathrm{TS}^2 \equiv \frac{\lambda^2}{(v_3^2 e)^2} (\widetilde{R}'_\mathrm{TS})^2\,.
\end{equation}
Note that in this case we keep the factors of $v_3$ outside of the integral to make it dimensionless, since the original radius $R^2_\mathrm{TS}$ is dimensionful.

\begin{figure}[t]
    \centering \includegraphics[width=0.5\linewidth]{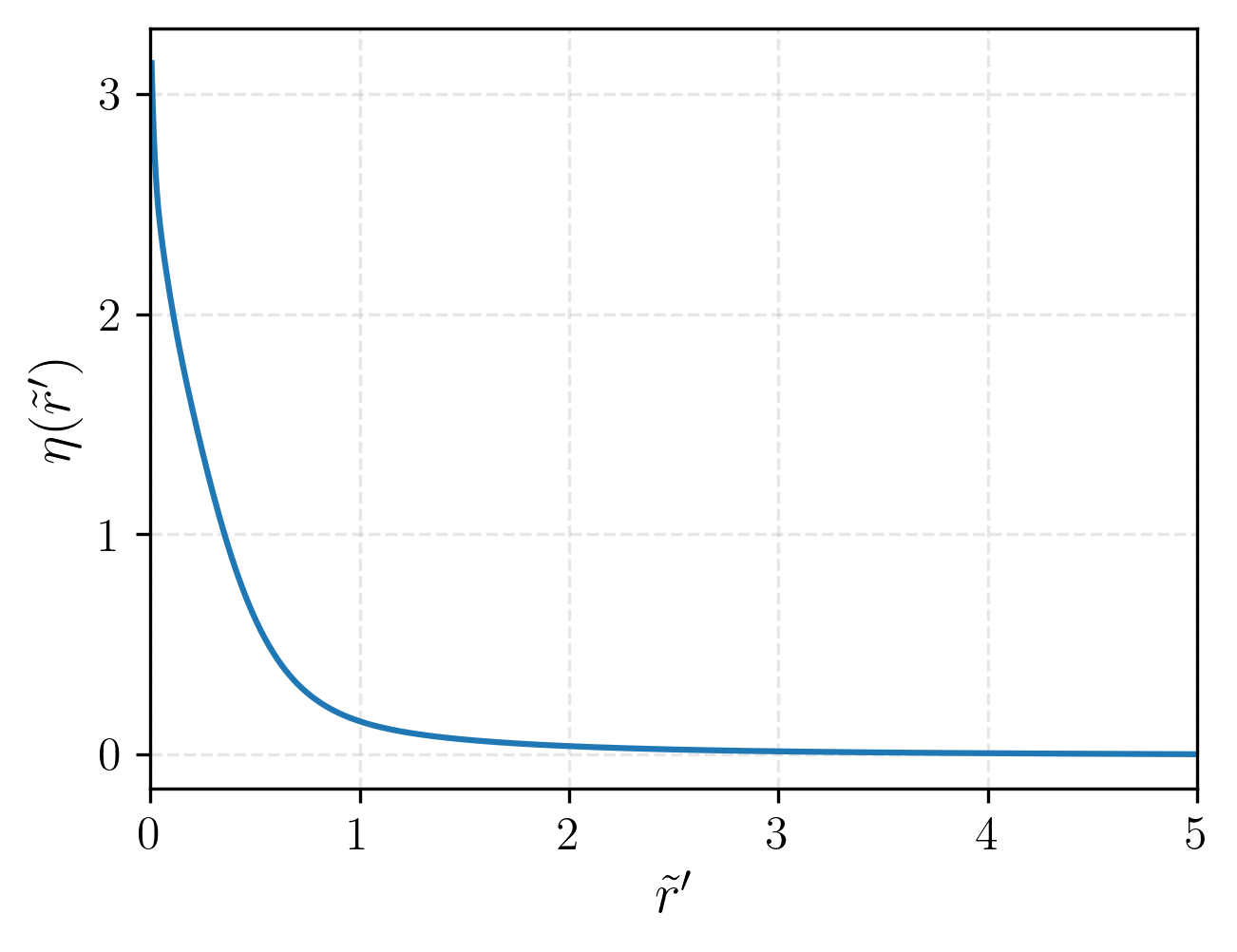}
    \caption{\it Thermoskyrmion profile that minimizes the rescaled 3D action $\widetilde{S}'_3[\eta]$ subject to boundary conditions $\eta(0)=\pi$ and $\eta(\rr'_\mathrm{max}=5)=0$; see text for details.}
    \label{fig:profile}
\end{figure}

Following this approach, we minimize the loss functional \eqref{eq:loss} for the rescaled action $\widetilde{S}'_3$ with $e=0.1$, and the resulting profile is shown in Figure \ref{fig:profile}. This configuration, which is reminiscent of the usual skyrmion profile (see Figure 1 in \cite{Adkins:1983ya}), perfectly satisfies the imposed boundary conditions ($n_U = 1$), minimizes the action and satisfies Derrick's criterion for stability, i.e. $\Delta \lesssim 10^{-2}$. Therefore, we can conclude that we found a stable classical configuration with unit winding number in the high-temperature limit of model B: a TS. Its action and size are~\footnote{The choice of the number of decimals in these results is arbitrary, as we show them without any measure of uncertainty. One could however roughly estimate the uncertainty as the deviation from the ideal $\Delta=0$ for stable configurations.}
\begin{equation}\label{eq:S R offshell}
    \widetilde{S}'_3[\eta] \thickapprox 0.970\,, \quad (\widetilde{R}'_\mathrm{TS})^2 \thickapprox 0.267\,.
\end{equation}

\subsection{Validity of the 3D EFT expansion}
\label{sec:validity}

The fact that the stabilization of TS implies a balance between two-derivative and six-derivative operators calls for an inspection of the validity of the 3D EFT expansion. Naively, the effective expansion is an expansion in derivatives over the heavy thermal modes scale, $\partial/(\pi T)$, so large field gradients might spoil its convergence.

Performing the full eight-derivative matching is beyond the scope of this project, but field redefinitions can be used to make an estimation of the missing orders. Indeed, the six-derivative off-shell terms in Eq. \eqref{eq:lagp6} can be redefined perturbatively in favor of a Lagrangian with the on-shell operators only. As explained in \cite{Scherer:1994wi}, these field redefinitions do not only remove the off-shell terms, but they also generate additional higher-derivative operators (which are usually neglected in a truncated EFT expansion). Therefore, by comparing our results in the off-shell and on-shell bases at six-derivatives, one can have a rough estimate of the size of the eight-derivative operators we have not included. If we carry out the minimization of the 3D action in the on-shell basis (setting the terms in gray to zero) we again find a TS configuration, with values
\begin{equation}\label{eq:S R onshell}
    \widetilde{S}'_3[\eta] \thickapprox 0.954\,, \quad (\widetilde{R}'_\mathrm{TS})^2 \thickapprox 0.259 \qquad \text{(on-shell basis)}\,,
\end{equation}
which differ from the ones presented before (off-shell basis, Eq. \eqref{eq:S R offshell}) only at the 1\% level. We can therefore conclude that truncating the 3D EFT at six-derivative level is justified, and that the TS are not an artifact of this truncation. 

\begin{figure}[t]
    \centering
    \includegraphics[width=0.5\linewidth]{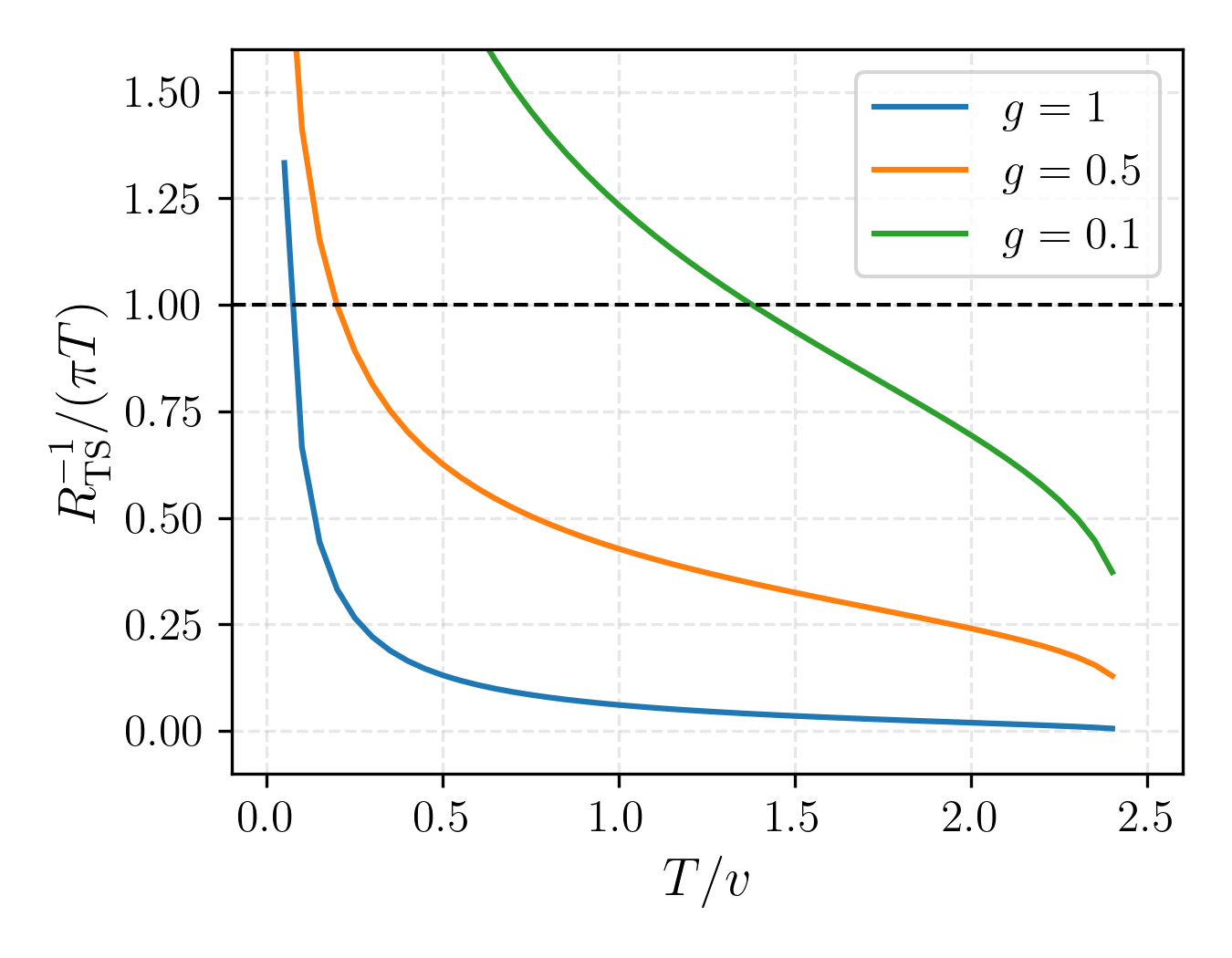}
    \caption{\it Ratio between the inverse thermoskyrmion radius $R_\mathrm{TS}^{-1}$ and the thermal scale $\pi T$ (in units of $v$) for $g=1, 0.5, 0.1$ in a range of temperatures. The dashed line marks a very conservative estimate of the breaking of the 3D EFT expansion on the low-temperature side, neglecting the suppression of operators due to their loop-generated WCs.}
    \label{fig:partialoverT}
\end{figure}

Alternatively, the convergence of the gradient expansion in $\partial/(\pi T)$ can be tested by approximating $\partial \sim R^{-1}_\mathrm{TS}$, so that the TS size cannot be too small compared to the heavy Matsubara scale. We show this ratio in Figure \ref{fig:partialoverT}, where for $g=1$ it is clear that temperature can reach fairly low values compared to the $v$ scale within the regime of validity of the 3D EFT. The situation appears less promising for $g=0.1$, as the expansion apparently breaks down at $T > v$.

Let us note that the plot in Figure \ref{fig:partialoverT} can be misleading at first glance, as we only compare gradients against the heavy thermal scale. In practice, however, there is an additional suppression due to the fact that all operators are loop generated in the matching, so operators with more derivatives and more field insertions come with larger powers of (perturbative) couplings and sum-integrals over more propagators (which are numerically smaller and which yield larger powers of $v/T$). If this additional suppression is taken into account, the effective expansion remains valid up to fairly low temperatures below $v$ for different perturbative values of $g$.

On the other hand, there also exists a natural cutoff at $\Lambda = 4 \pi v$, beyond which the effective expansion of non-linear field theories becomes invalid \cite{Manohar:1983md}. This is evident in the loop expansion of the 3D kinetic term (Eq. \eqref{eq:c0 model B}), where the ratio $T^2/v^2$ implies that $T$ cannot take on too-high values (in fact, $T < \sqrt{6} v$ in order for this term to be positive). Let us note that although we call it a \textit{high-temperature expansion} (in derivatives over temperature) in this work, it is also a \textit{low-temperature expansion} (with respect to the cut-off $\Lambda = 4 \pi v$), as first coined in the context of $\chi$PT at finite temperature \cite{Gasser:1986vb}.

All in all, schematically, our construction is valid in the regime 
\begin{equation}\label{eq:T range}
    \frac{1}{R_\mathrm{TS}} \lesssim \pi T \lesssim 4 \pi v\,,
\end{equation}  
where the lower limit can be actually pushed further down due to the numerical suppression of higher-derivative operators by their WCs~\footnote{One could also wonder whether 2-loop matching corrections to four-derivative terms dominate over the 1-loop corrections to six-derivative operators that we have computed. This is not the case. Indeed, on the basis of diagrams topologies and dimensional analysis, it can be seen that $c_{1,2}^{\text{2-loop}}\sim g^2 T^2/v^2$ while $c_{6}^\text{1-loop}\sim g^2$, from where $$\frac{1}{T^2}\frac{c_{1,2}^{\text{2-loop}}\partial^4}{c_{6}^\text{1-loop}\partial^6}\sim \frac{T^4}{v^2\partial^2}\sim \frac{T^4 R_\text{TS}^2}{v^2}\,,$$ where, as before, we have estimated $\partial\sim R_\text{TS}^{-1}$. For $g=1, 0.5, 0.1$, this ratio is below $1$ for values $T/v < 0.9, 1.2, 2.1$, respectively. Naturally, this corresponds to the region of parameter space where we have previously shown that the EFT expansion shows a better convergence; see Eq. \eqref{eq:T range}.}.

\subsection{Thermodynamics}

In model B we find stable TS configurations $U_\mathrm{TS}(\mathbf{x})$ which minimize the rescaled action $\widetilde{S}'_3[\eta]$ with unit winding number. Recovering the original units, we have then solved
\begin{equation}
    \frac{\delta S_3}{\delta U}\biggr|_{U=U_\mathrm{TS}} = 0
\end{equation}
in the hedgehog ansatz and with appropriate boundary conditions.

In the saddle-point approximation, if fluctuations of the zero Matsubara modes on the TS background are neglected, the partition function (generating functional) for a single TS can be approximated as
\begin{equation}
    \mathcal{Z} \thickapprox e^{-S_3[U_\mathrm{TS}]}\,.
\end{equation}
Now, from the partition function, the thermodynamic free energy $F$, entropy $S$ and average energy $E$ of the system are defined as
\begin{align}
    F &= - T \log \mathcal{Z}\,, \\
    S &= -\frac{\partial F}{\partial T}\,, \\
    E &= F + T S = F - T \frac{\partial F}{\partial T}\,.
\end{align}
Using the
approximated partition function for the single TS,
the values for the thermodynamic magnitudes are given by 
\begin{align}
    F &\thickapprox T S_3[U_\mathrm{TS}]\,, \\
    S &\thickapprox -S_3[U_\mathrm{TS}] - T \frac{\partial S_3[U_\mathrm{TS}]}{\partial T}\,,\\
    E &\thickapprox - T^2 \frac{\partial S_3[U_\mathrm{TS}]}{\partial T}\,.
\end{align}
Therefore, the TS can be interpreted as an extended configuration of the fields that minimizes the free energy $F$, instead of the energy, as is the case for $T=0$ skyrmions. Also, since Derrick's criterion is satisfied, these solutions are stable against thermal fluctuations. 

\begin{figure}[t]
    \centering
    \includegraphics[width=0.325\textwidth]{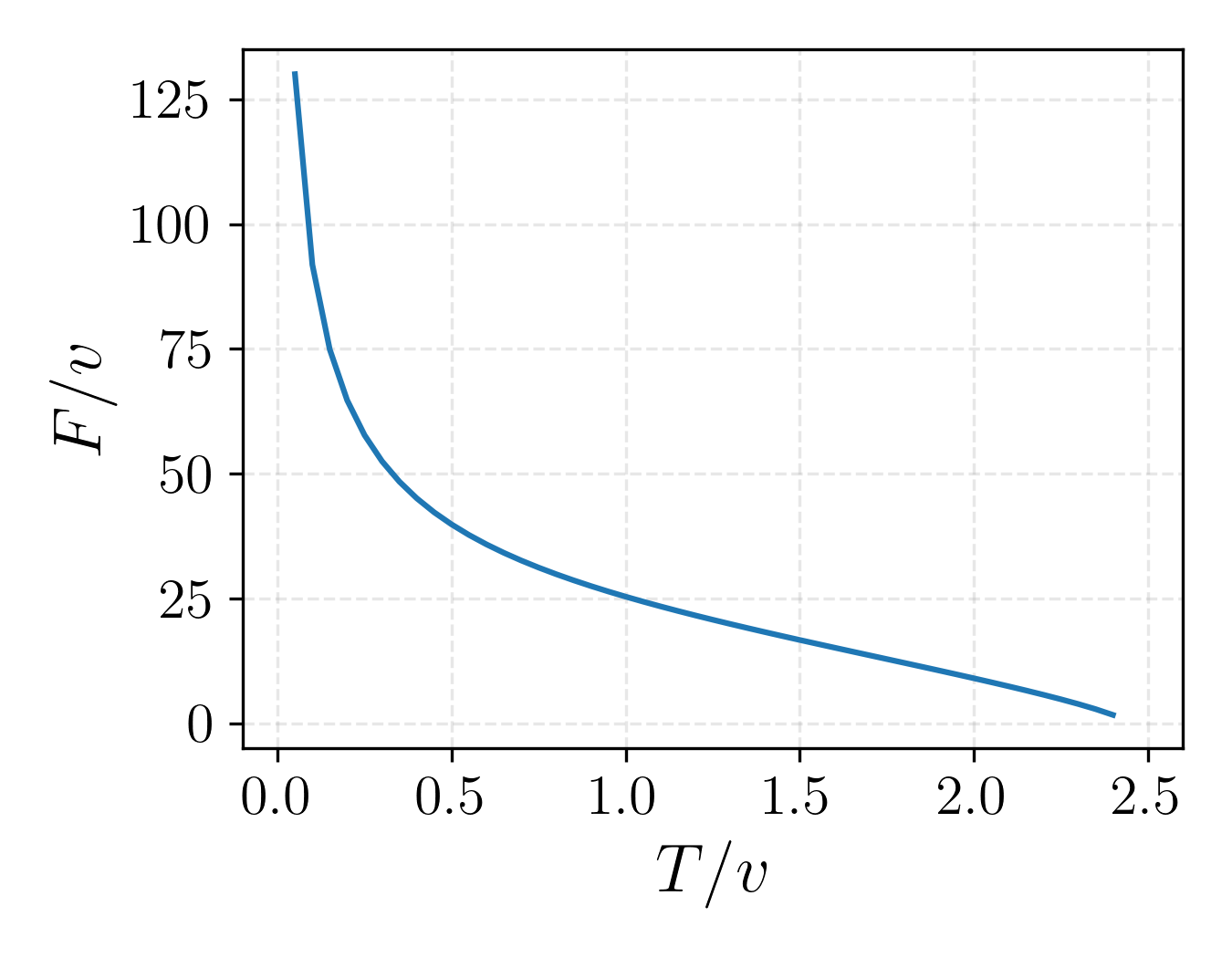}
    \includegraphics[width=0.325\textwidth]{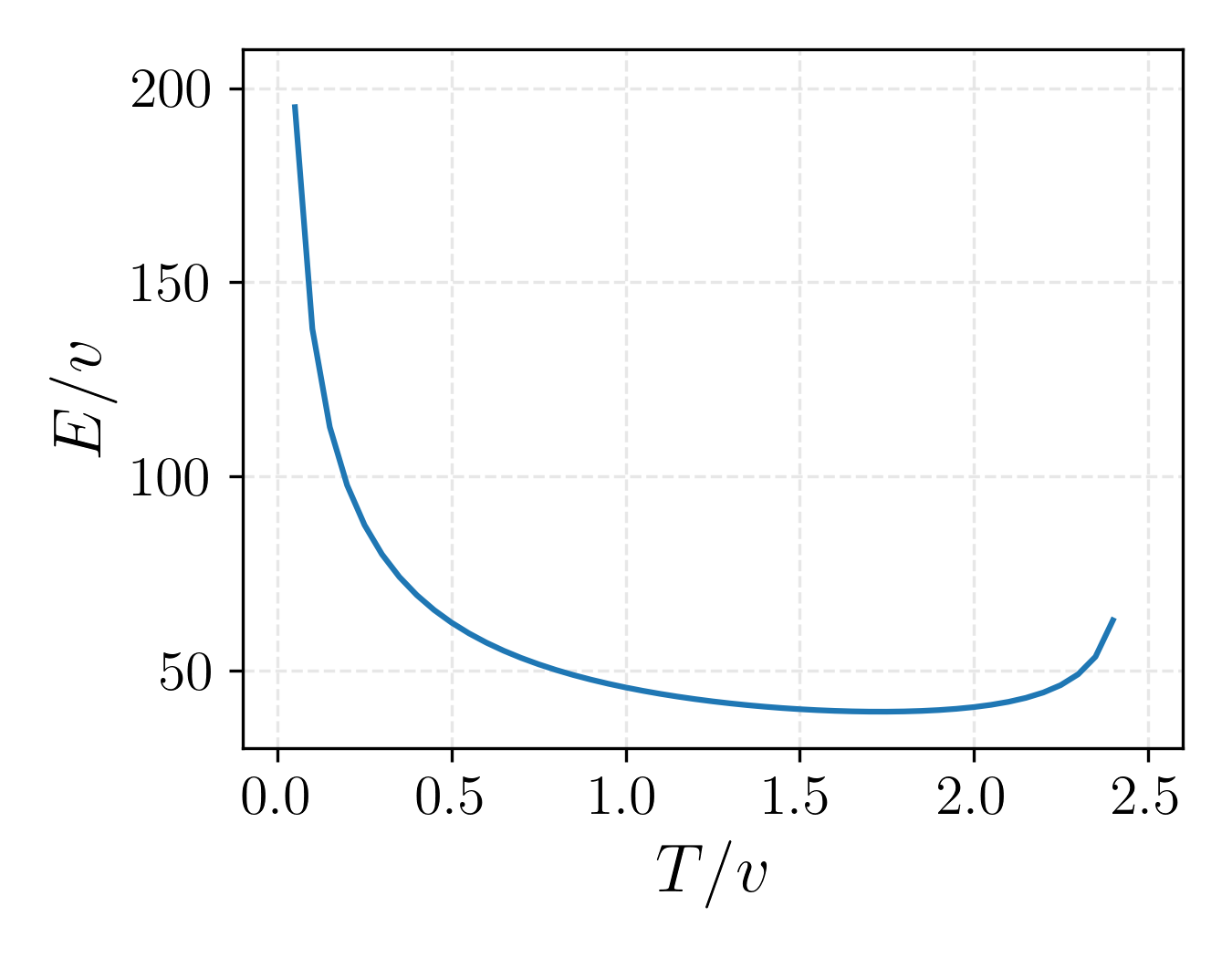}
    \includegraphics[width=0.325\textwidth]{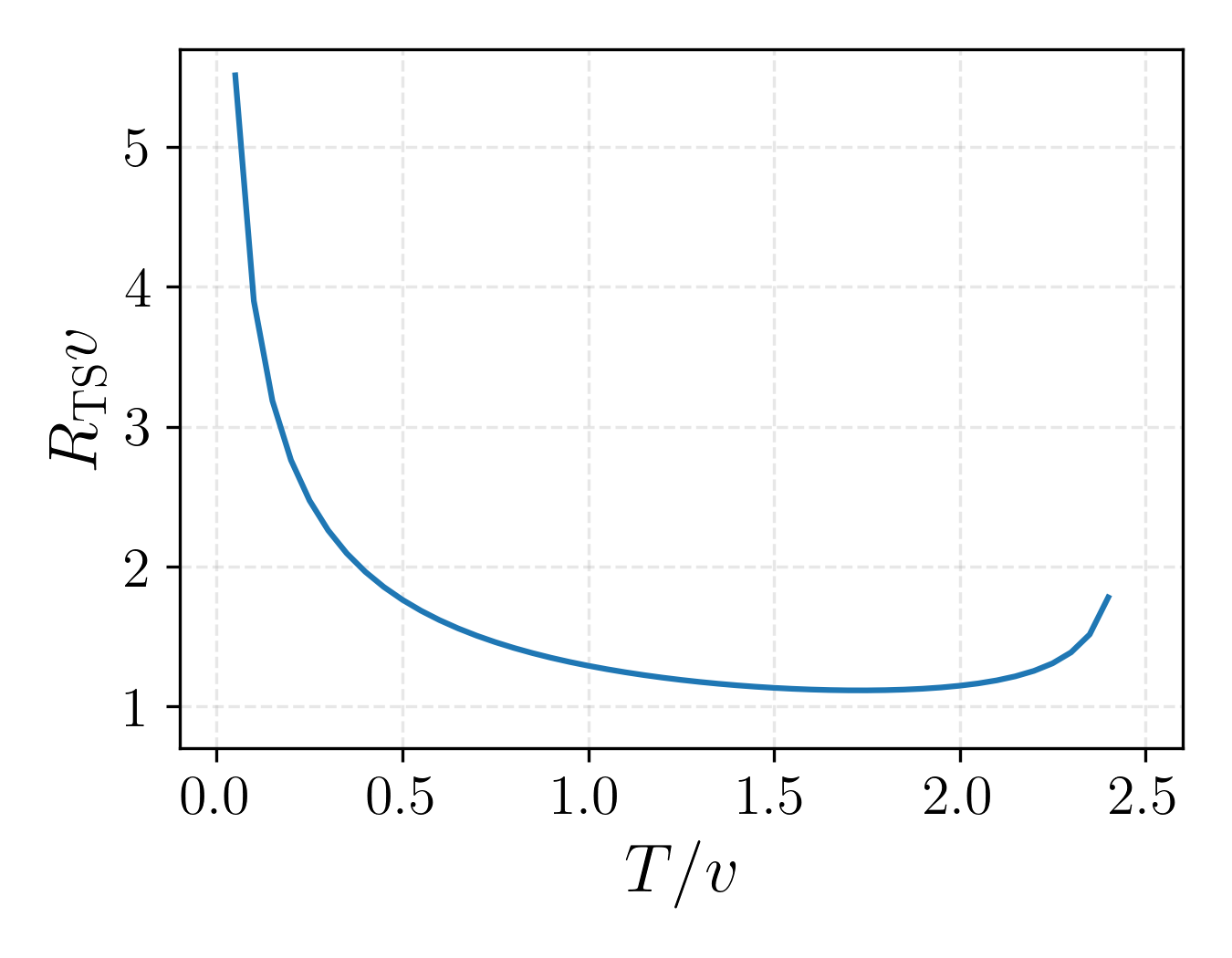}
    \caption{\it Free energy (top left), average energy (top right) and radius (bottom) of a thermoskyrmion in units of $v$ for fixed $g=1$, in the allowed temperature range.}
    \label{fig:energyandradius}
\end{figure}

Reinserting the rescaling factors, in Figure \ref{fig:energyandradius} we show the free energy, average energy and radius of the TS
in a range of temperatures. The EFT expansion 
breaks down at around $T/v \sim \sqrt{6}$, a value for $T$ beyond which $c_0$ becomes negative (Eq. \eqref{eq:c0 model B}) and thus the scaling factor $\lambda$ acquires an imaginary part. At lower temperatures, the three magnitudes follow a $T^{-1/2}$ scaling, which can be understood from their expressions in a low-temperature approximation. We see that for decreasing temperatures, the three magnitudes increase following this power law until the limit of validity of the high-temperature description, $T/v \sim 0.1$, is reached (see Figure \ref{fig:partialoverT}).

It can be seen that, as the system adiabatically cools down, the TS size grows rapidly and so do the free and average energies. Furthermore, estimating the TS mass by the average energy $E$, the TS is always much heavier than the temperature, thus making it a non-relativistic species. It is also one or two orders of magnitude larger than the $v$ scale. 

\subsection{A note about the classical treatment}

TS, just like regular skyrmions, are classical configurations of a non-linear field theory. In our work, the necessary stabilizing, higher-derivative operators appear as an effective description of quantum (loop) fermionic interactions which are integrated out. In this sense, one could claim that TS are \textit{semiclassical}, as the bosonic zero Matsubara modes in $U$ are effectively corrected by quantum effects of all heavy thermal modes.

\begin{figure}
    \centering
    \includegraphics[width=0.5\textwidth]{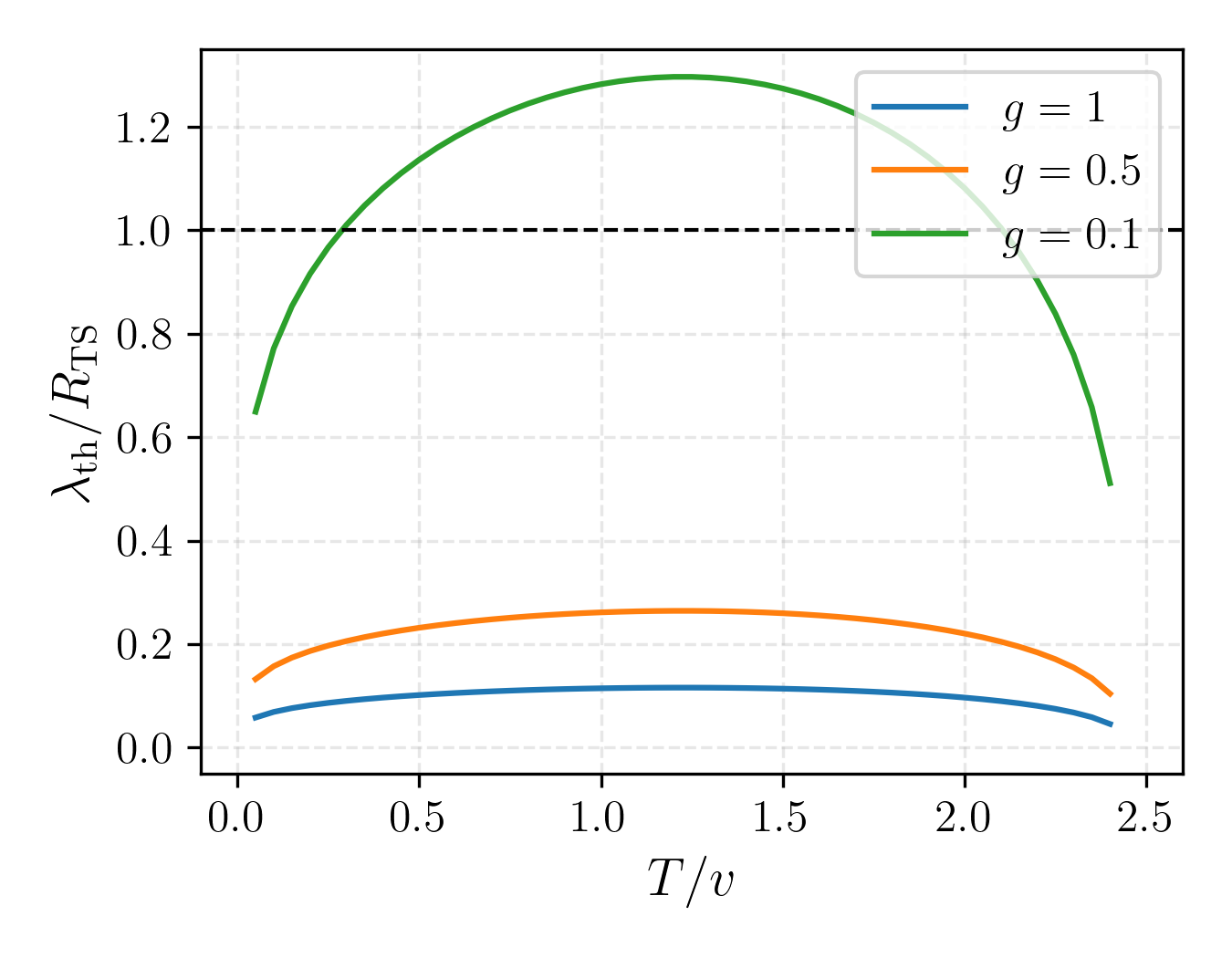}
    \caption{\it Ratio $\lambda_\mathrm{th}/R_\mathrm{TS}$ for $g=1,0.5,0.1$, in the allowed temperature range. The dashed line marks the breaking of the assumption of classicality for the thermoskyrmion.}
    \label{fig:debroglie}
\end{figure}

Although quantum corrections are partially accounted for, one is left to wonder if the treatment of TS as approximately classical objects (a particle-like object with definite position and size) is justified in the thermodynamical discussion above. One can have an estimate of the \textit{quantum fuzziness} of a particle in an ideal gas by comparing its size to the characteristic size of quantum thermal fluctuations. This scale is given by the thermal de Broglie wavelength \cite{Zijun:2000},
\begin{equation}
    \lambda_\mathrm{th} \sim \frac{1}{\sqrt{T E}}\,,
\end{equation}
where we again estimate the TS mass by $E$. In order for the TS to be a classical object, its radius must be larger than this wavelength, which we inspect in a range of temperatures in Figure \ref{fig:debroglie}. We see that for $g=1, 0.5$ the ratio between the two remains small throughout, the size being one order of magnitude greater than the wavelength. With $g=0.1$, quantum effects seem to dominate for a wide range of temperatures, but in most of this range the EFT expansion is not valid (see Figure \ref{fig:partialoverT}), so we cannot make statements about the nature or existence of TS.

This rough check reveals that TS can approximately behave as classical objects for $g>0.1$. In any case a more detailed study of the quantum corrections and properties should still be performed. We relegate this analysis to future work. 

\subsection{The role of the temperature}
\label{sec:T role}

The appearance of skyrmions in the high-temperature regime of model B might at first seem surprising. At $T=0$, the model contains fermions, which cannot take on a classical field configurations; also, there are no higher-derivative stabilizing operators for the $U$ field, so classical skyrmion configurations do not exist. However, in a thermal background, the hierarchy of scales between thermal modes allows one to construct a local effective action for the zero mode of $U$, where stabilizing operators do appear. In this setup, we have found classical skyrmion configurations for the zero mode of $U$. 

It is relevant to emphasize that temperature might not play any dynamical role in stabilizing the skyrmions, but rather enable a regime in which a controlled, local semiclassical description of the bosonic sector can be constructed. The operators which stabilize the skyrmion solutions are an effective description of fermion-boson interactions in a local expansion in derivatives over temperature. In this situation, the analysis of skyrmion solutions proceeds in precisely the same manner as in classical field theory.

The situation at zero temperature is somewhat different. There, the presence of light fermions obstructs the construction of a purely bosonic local effective action, rendering the question of skyrmion existence harder to address. This difficulty implies that skyrmion-like configurations must be stabilized through quantum effects, but it does not impede their existence. In fact, the thermal picture ---which, though differently, accounts for certain quantum corrections--- suggests that they might be also present at zero temperature. Were this to happen in the electroweak sector of the SM, the skyrmions would be massive, neutral and stable, and so potentially dark matter candidates.

\section{Conclusions}
\label{sec:conclusions}

We have explored for the first time the skyrmion configurations that arise in quantum field theory (QFT) at finite temperature ($T$), using a $SU(2)$ symmetric toy model coupling a non-linear scalar $U$ to fermions $\psi$. Inspired by techniques developed for hot QCD and cosmological phase transitions, the hierarchy of thermal modes has allowed us to construct a 3-dimensional effective field theory (3D EFT) for the bosonic field $U$ in a derivative expansion, analogous to the pion chiral Lagrangian. Employing a neural network minimization algorithm, we have found stable, topological configurations that minimize the 3D EFT action and dubbed them thermoskyrmions (TS).

In this thermal setup, higher-derivative operators, which are the key to stability by virtue of Derrick's theorem, are generated in a predictive, perturbative matching procedure by integrating out heavy thermal modes. In contrast with previous works, here the necessary higher-derivative operators are neither (a) generated due to confinement at a larger scale, as is the case with the low-energy descriptions of QCD or composite Higgs models, nor (b) put by hand in an EFT (like SMEFT or HEFT), invoking some UV dynamics that generate them with the appropriate relative signs. In our toy model, higher-derivative operators are simply the effective description of thermal fermionic interactions and appear as local operators due to an EFT expansion in gradients over temperature.


Even though stability requires that two- and higher-derivative operators be of similar size, we have seen that the effective expansion converges and can be truncated in a well-defined temperature range. The key is that higher-derivative operators are loop-generated, and thus come with increasingly large numerical suppression. This supports the idea that strongly-coupled dynamics is not essential to generate large hierarchies between different higher-derivative operators, as miraculously happens in the pion chiral Lagrangian with the four-derivative Skyrme operator.

We have interpreted TS as topologically-protected, collective excitations of the fields in a thermal medium which are stable minima of the free energy. As such, they are different from the \textit{bounce} configuration, which is an unstable minimum of the free energy and describes the most likely path for thermal vacuum decay. In our toy model, the average energy (mass) of TS increases as the system cools down, and it can reach values which are one or two orders of magnitude larger than the temperature scale, which make them non-relativistic species. The phenomenological implications of these particle-like configurations are still unknown to us. Were they to appear in more physically motivated models~\footnote{
Even though preliminary, we have some evidence that, in the presence of both $y$ and $g$ couplings ---namely in a better approximation of the SM---, TS-stabilizing operators can appear already at $\mathcal{O}(p^4)$.
}
, they might constitute a dark matter candidate without the need for new degrees of freedom, in line with \cite{Kitano:2016ooc,Murayama:2009nj}. It would also be interesting to explore if they can be realized in condensed matter systems \cite{Lancaster03072019}.


Finally, we would like to remark that there is no evidence that analogous stable configurations cannot exist at zero temperature. In fact, their existence at finite temperature raises the question as to how these configurations smoothly evolve towards the zero-temperature limit, beyond the reach of the 3D EFT construction. As we see it, the main effect of the temperature is to act as a scale over which the quantum effective action for bosons can be perturbatively expanded in local operators. In the absence of such scale (at $T=0$), our toy model does not show any mass hierarchy, and since it contains fermions and no higher-derivative operators for the $U$, skyrmions cannot exist, at least, as classical static configurations that minimize the action. At zero temperature, one should then construct a bosonic quantum effective action without relying on any local expansion and numerically determine if such configurations exist. In such case, these skyrmions would be intrinsically quantum and would not require higher-derivative operators in the classical action, opening up completely unexplored avenues in the study of topological configurations. In particular, our findings point at the electroweak sector with fermions (even at renormalizable level) after electroweak symmetry breaking as a potential candidate to hold such skyrmionic configurations. This constitutes a promising line of future work.

\section*{Acknowledgments}
We thank Javier L\'opez-Miras for help cross-checking a few matching results with (a yet-not-public version of) \texttt{matchete}~\cite{Fuentes-Martin:2022jrf}. LG is indebted to Oliver Gould and Farbod-Sayyed Rassouli for useful discussions and their hospitality during his visit to University of Nottingham. MC is supported by the European Research Council under grant agreement n. 101230200.  JCC is supported by the Ram\'on y Cajal program under grant number RYC2021-030842-I. LG is supported by the FPU program under grant number FPU23/02026. This work has received further funding from MICIU/AEI/10.13039/ 501100011033 (grants PID2022-139466NB-C21/C22 and PID2024-161668NB-100) as well as from Junta de Andaluc\'ia (grants FQM 101 and P21-00199).

\appendix

\section{Dimensional reduction in a non-linear field theory}
\label{app:matching}
In this appendix we describe how to construct a 3D EFT for a non-linear field theory through the process of high-temperature DR. In particular, we will explain how to determine the WCs in the 3D EFT via off-shell matching, and the useful consistency relations that arise from the non-linear nature of the theory.

\subsection{Setup}

As briefly introduced in subsection \ref{sec:tft}, we work in the imaginary time formalism, where the physics of thermal equilibrium is studied in a QFT where the time direction is compactified to a circle of length $2 \pi T$. In this spacetime, fields are Fourier decomposed into towers of Matsubara modes with different associated frequencies, which are different for bosons and fermions \cite{Matsubara:1955ws}.

Since the generating functional integrates over periodic configurations of the Goldstone matrix $U$, one could naively assume that
\begin{equation}
    U(\tau, \mathbf{x}) = \sum_{n=-\infty}^\infty U_n(\mathbf{x}) e^{i \omega_n \tau}\,,
\end{equation}
and the goal would be to construct the 3D EFT for $U_0(\mathbf{x})$. However, this decomposition spoils the unitarity of $U$, because in general
\begin{equation}
    U^\dagger U = \sum_{n=-\infty}^\infty \sum_{m=-\infty}^\infty U_n^\dagger(\mathbf{x}) U_m(\mathbf{x}) e^{i (\omega_m - \omega_n) \tau} \neq \mathbb{1}_2\,,
\end{equation}
and in particular the zero mode would not be unitary. The correct approach is to decompose the three Goldstone components $\pi_a$, and this naturally preserves the unitarity of their non-linear representation. Indeed, if the decay constant $v$ is absorbed in the $\pi_a$, then:
\begin{equation}
    U(\tau, \mathbf{x}) \equiv \exp \left\{i \pi_{a}(\tau, \mathbf{x}) \sigma_a \right\} = \exp \left\{i \sum_{n=-\infty}^\infty \pi_{a,n}(\mathbf{x}) \sigma_a e^{i \omega_n \tau} \right\} \,;
\end{equation}
hence the reality condition $(\pi_{a,n})^* = \pi_{a,-n}$ \cite{Laine:2016hma} and the fact that $\omega_{-n} = -\omega_n$ imply
\begin{equation}
    U^\dagger(\tau, \mathbf{x}) = \exp \left\{-i \sum_{n=-\infty}^\infty \pi_{a,-n}(\mathbf{x}) \sigma_a e^{i \omega_{-n} \tau} \right\} = \exp \left\{-i \sum_{n=-\infty}^\infty \pi_{a,n}(\mathbf{x}) \sigma_a e^{i \omega_{n} \tau} \right\}\,,
\end{equation}
from where it is evident that $U^\dagger U = \mathbb{1}_2$, and also each mode is represented by an unitary matrix. 

Reintroducing the decay constant, the temporal part of the kinetic term at finite temperature reads
\begin{equation}
    \mathcal{L} \supset \frac{v^2}{4} \trace{\partial_0 U^\dagger \partial_0 U} = \frac{1}{2} \sum_{n=-\infty}^\infty \omega_n^2 (\pi_{a,n})^2 + \mathcal{O}(\pi^4)\,,
\end{equation}
so non-zero modes of the $\pi_a$ acquire a thermal mass, and can thus be perturbatively integrated out. The same happens for fermions $\psi$, but for them all modes acquire a thermal mass and are therefore absent in the low-energy theory. The relevant degree of freedom in the 3D EFT is
the bosonic zero mode matrix $U_0(\mathbf{x}) \equiv \exp \left\{i \pi_{a,0}(\mathbf{x}) \sigma_a / v_3 \right\}$, where $v_3 \equiv v / \sqrt{T}$. Following the standard convention in DR, the fields $\pi_a$ are normalized by factors of $T$ so that they have units of $\sqrt{T}$ and the 3D action be dimensionless. In what follows, we drop the zero subscript from the zero modes in the 3D EFT to ease the notation.

In order to determine the 3D EFT, we match off-shell correlators with only the light, zero modes of the $\pi_a$ in external legs. Since matching is performed at the level of the fields $\pi_a$, $U$ must be expanded to obtain all possible vertices that can contribute to each $n$-point function. While truncating this expansion at any given order breaks the unitarity of $U$, accounting for all terms that contribute to a given $n$-point function is sufficient to compute it exactly (up to a given loop order). This expansion of $U$ is carried out both in the 4D theory and in the 3D theory, so in practice the matching is performed at the level of the $\pi_a$. 

Different operators involving the $U$ lead to different kinematic structures for the $\pi_a$ in the Feynman rules, so one can easily compute the 3D EFT of a non-linear field theory by computing its linear expansion up to high-enough order, matching for the WCs and then writing the 3D EFT back in terms of operators in the non-linear representation. Furthermore, the fact that an operator with $n$ insertions of $U$ in the 3D EFT can be expanded to yield vertices with an arbitrary number $m \geq n$ of $\pi_a$, and thus contribute to any Green's function with more than $n$ external legs, poses powerful consistency constrains on the matching.

Since a good example is worth a thousand words, let us illustrate this 
procedure with a particular matching computation in model A.

\subsection{Example: Matching in model A}

The 4D Lagrangian of model A in Minkowski spacetime reads:
\begin{equation}\label{eq:UVlag A}
    \mathcal{L} = \frac{v^2}{4} \trace{\partial_\mu U^\dagger \partial^\mu U} + i \overline{\psi}_L \gamma_\mu \partial^\mu \psi_L + i \overline{\psi}_R \gamma_\mu \partial^\mu \psi_R - \frac{y v}{\sqrt{2}} \left( \overline{\psi}_L U \psi_R + \text{h.c.} \right) \,,
\end{equation}
and the most general 3D EFT Lagrangian up to $\mathcal{O}(p^4)$ is given in Eqs. \eqref{eq:lagp2} and \eqref{eq:lagp4}.

Our goal is to first compute the kinetic term in the 3D EFT, so we must determine and solve the matching equation for the coefficient $c_0$. We do this by equating the off-shell 2-point function for $\pi_a$, with $a=1,2,3$, between the 4D and the 3D theories, and we do it at 1-loop order. We employ a routine based on \texttt{FeynRules} \cite{Alloul:2013bka} to compute the Euclidean Feynman rules, and we compute the relevant diagrams using \texttt{FeynArts} \cite{Hahn:2000kx} and \texttt{FeynCalc} \cite{Shtabovenko:2023idz}.

The tree-level part in Euclidean space trivially reads 
\begin{equation}
    \Gamma^\text{4D, tree}_{\pi_a \to \pi_a} = - |\mathbf{p}|^2\,,
\end{equation}
where $P = (0, \mathbf{p})$ is the Euclidean external momentum of the zero modes.

\begin{figure*}[t]
    \centering
    \includegraphics[width=0.8\textwidth]{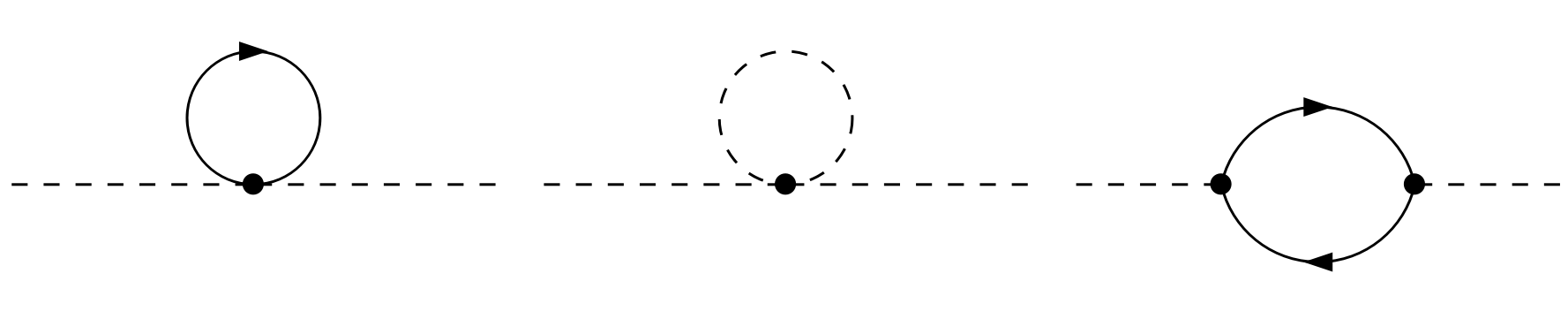}
    \caption{\it Diagrams contributing to the two-point function of $\pi_a$ at 1-loop order.}
    \label{fig:diagrams}
\end{figure*}

At 1-loop, any diagram contributing to the 2-point function contains at most vertices with four fields, so we expand the $U$ in Eq. \eqref{eq:UVlag A} to find all $(\pi_a)^4$ and $\psi^2 (\pi_a)^2$ vertices. From there, we derive the relevant Feynman rules and the 1-loop contribution reads:
\begin{equation}
    \Gamma^\text{4D, 1-loop}_{\pi_a \to \pi_a} = 4 y^2 \sumintF{Q} \left[ \frac{m_\psi^2 + Q^2 + P \cdot Q}{(Q^2 + m_\psi^2) ((Q-P)^2 + m_\psi^2)} - \frac{1}{Q^2 + m_\psi^2} \right] + \frac{2 P^2}{3 v^2} \sumintB{Q} \frac{1}{Q^2}\,,
\end{equation}
where in the sum-integrals the $Q$ ($\{Q\}$) subscript denotes a sum over bosonic (fermionic) frequencies, and $Q$ is the Euclidean loop momentum. In Figure \ref{fig:diagrams} we show the relevant diagrams.

Now we perform a hard-region expansion \cite{Beneke:1997zp}, defined by $Q^2, (\pi T)^2 \gg  P^2, m_\psi^2$, which yields a series of terms like the ones shown in Eq. \eqref{eq:hrexpansion}. We truncate the mass expansion at $\mathcal{O}(y^4)$, and to achieve $\mathcal{O}(p^2)$ (relevant order for the 3D kinetic term), we iterate the following algebraic identity:
\begin{equation}
 \frac{1}{(Q+P)^2} = \frac{1}{Q^2}\left[1-\frac{P^2+2Q\cdot P}{(Q+P)^2}\right]\,.
\end{equation}
Following some algebraic manipulations, this allows us to write the 2-point function as a sum of massless, tadpole sum-integrals. At 1-loop order, all bosonic sum-integrals are given by the following master formula in the $\overline{\mathrm{MS}}$ scheme in dimensional regularization:
\begin{equation}
    \hat{I}_{\alpha}^{r} \equiv \sumintB{Q} \frac{Q_0^r}{Q^{2 \alpha}} = \tilde{\mu}^{2\epsilon} \frac{\left( 1 + (-1)^r \right) T}{(2 \pi T)^{2 \alpha - r - d}} \frac{\Gamma\left( \alpha - d/2 \right)}{(4 \pi)^{d/2} \Gamma\left( \alpha \right)} \zeta\left( 2 \alpha - r - d \right) \,,
\end{equation}
where $d=3-2\epsilon$, $\Gamma(x)$ is the Euler gamma function and $\zeta(x)$ is the Riemann zeta function. On the other hand, the fermionic sum-integrals are simply given by:
\begin{equation}
    I_{\alpha}^{r} \equiv \sumintF{Q} \frac{Q_0^r}{Q^{2 \alpha}} = \left( 2^{2\alpha - r - d} - 1 \right) \hat{I}_{\alpha}^{r}\,.
\end{equation}

Applying all of the above, the result in the 4D side reads:
\begin{equation}
    \Gamma^\text{1-loop}_{\pi_a \to \pi_a} = \left( \frac{T^2}{18 v^2} + \frac{7 \zeta(3) y^4}{(4 \pi)^4} \frac{v^2}{T^2} \right) |\mathbf{p}|^2\,.
\end{equation}
Let us note that there is no momentumless correction to the 2-point function, which is consistent with the fact that the zero modes of $\pi_a$ are massless. 

Now, in the 3D EFT, the same 2-point function in the hard-region expansion is simply the tree-level contribution, so, at $\mathcal{O}(p^2)$,
\begin{equation}
    \Gamma^\text{3D, tree}_{\pi_a \to \pi_a} = - 4 c_0 \frac{T^2}{v^2} |\mathbf{p}|^2\,.
\end{equation}

Therefore, the matching equation is:
\begin{equation}
    \Gamma^\text{3D, tree}_{\pi_a \to \pi_a} = \Gamma^\text{4D, tree}_{\pi_a \to \pi_a} + \Gamma^\text{4D, 1-loop}_{\pi_a \to \pi_a} \Longrightarrow c_0 = \frac{v^2}{4 T^2} \left( 1 - \frac{T^2}{6 v^2} - \frac{28 \zeta(3) y^4}{(4 \pi)^4} \frac{v^2}{T^2} \right)\,,
\end{equation}
as shown in Eq. \eqref{eq:modelA c0} in the main text.

The same reasoning can be followed to match the $c_3$ operator from the 2-point function, since it also contributes to the two-$\pi$ vertex at four-derivative level, but the hard-region expansion must be extended to $\mathcal{O}(p^4)$.

Next, 
to compute the matching to the other four-derivative operators in the 3D EFT, we 
notice that $c_1$ and $c_2$ contribute to the 4-point function (with distinct kinematic structures) but not to the 2-point. Diagrams with four legs at 1-loop can contain vertices with up to six fields, so now we 
expand $U$ in the Lagrangian up to $(\pi_a)^6$ and $\psi^2 (\pi_a)^4$ to compute a new set of Feynman rules for the 4-point function.

The other two operators, $c_0$ and $c_3$, contribute both to the 2-point and the 4-point. This means that by computing the 4-point function they can be determined independently, and the result must be consistent with that obtained from the matching of the 2-point function. This constitutes a powerful cross-check that the matching is correct at each order.

The full matching equations for the 4-point function are rather lengthy, as they involve several kinematic structures involving three independent external momenta, so we refrain from writing them here. Instead, let us show how the $\mathcal{O}(p^2)$ fermionic contribution yields the same matching equation to $c_0$ as in the 2-point function.

In the 3D EFT side, at tree-level, the $\pi_1 \pi_1 \to \pi_3 \pi_3$ function reads:
\begin{equation}
    \Gamma^\text{3D, tree}_{\pi_1 \pi_1 \to \pi_3 \pi_3} = \frac{4 c_0 T^2}{3 v^4} \left( |\mathbf{p}_1|^2 + 4 \mathbf{p}_1 \cdot \mathbf{p}_2 + 2 \mathbf{p}_1 \cdot \mathbf{p}_3 + |\mathbf{p}_2|^2 + 2 \mathbf{p}_2 \cdot \mathbf{p}_3 - 2 |\mathbf{p}_3|^2 \right) + \dots
\end{equation}
We must reproduce the same kinematic structure in the 4D theory at 1-loop to match the $c_0$ coefficient. As an example, we 
only show the $|\mathbf{p}_1|^2$ piece. After applying the hard-region expansion up to $\mathcal{O}(y^4)$ and tensor-reducing the integrals, it reads:
\begin{equation}
    \Gamma^\text{4D, 1-loop}_{\pi_1 \pi_1 \to \pi_3 \pi_3} = |\mathbf{p}_1|^2 \left[ \frac{4 y^2}{3 v^2} \frac{1}{d} \sumintF{Q} \left( \frac{2-d}{Q^4} - \frac{2 Q_0^2}{Q^6} \right) + \frac{4 y^4}{3} \frac{1}{d} \sumintF{Q} \left( \frac{d - 3}{Q^6} + \frac{3 Q_0^2}{Q^8} \right) \right] + \dots
\end{equation}
%

Replacing the sum-integrals by the master formulae above, the $\mathcal{O}(y^2)$ piece vanishes and we get:
\begin{equation}
    \Gamma^\text{4D, 1-loop}_{\pi_1 \pi_1 \to \pi_3 \pi_3} = \frac{7 \zeta(3) y^4}{192 \pi^4 T^2} |\mathbf{p}_1|^2 + \dots \,.
\end{equation}
Therefore, when equating the two we again get the fermionic contribution to $c_0$ shown in Eq. \eqref{eq:modelA c0}. 

\section{Effective operators at $\mathcal{O}(p^6)$}
\label{app:operators}

The full basis of $\mathcal{O}(p^6)$ operators in the chiral Lagrangian was first derived in \cite{Fearing:1994ga}. This appendix contains the spatial part of the operators that appear in Eq. \eqref{eq:lagp6}, as found in Tables II - VIII in the aforementioned work in the gaugeless limit.

They read:
\begin{align}
    \mathcal{O}_{53} &= \trace{[\partial_i \partial_j U]_{-} [\partial_i \partial_j U]_{-} [\partial_k U]_{-} [\partial_k U]_{-}} \,, \\
    \mathcal{O}_{54} &= \trace{[\partial_i \partial_j U]_{-} [\partial_i \partial_k U]_{-} [\partial_j U]_{-} [\partial_k U]_{-}} \,, \\
    \mathcal{O}_{55} &= \trace{[\partial_i \partial_j U]_{-} [\partial_i \partial_k U]_{-} [\partial_k U]_{-} [\partial_j U]_{-}} \,, \\
    \mathcal{O}_{58} &= \trace{[\partial_i \partial_j U]_{-} [\partial_k U]_{-}} \trace{ [\partial_i \partial_j U]_{-} [\partial_k U]_{-}} \,, \\
    \mathcal{O}_{100} &= \trace{[\partial_i U]_{-} [\partial_i U]_{-} [\partial_j U]_{-} [\partial_j U]_{-} [\partial_k U]_{-} [\partial_k U]_{-} } \,, \\
    \mathcal{O}_{101} &= \trace{[\partial_i U]_{-} [\partial_i U]_{-} [\partial_j U]_{-} [\partial_k U]_{-} [\partial_j U]_{-} [\partial_k U]_{-} } \,, \\
    \textcolor{gray}{\mathcal{O}_{2}} &\textcolor{gray}{= \trace{ \mathcal{O}_\mathrm{EOM}^{(2)} \mathcal{O}_\mathrm{EOM}^{(2)} [\partial_i U]_{-} [\partial_i U]_{-} }} \,, \\
    \textcolor{gray}{\mathcal{O}_{3}} &\textcolor{gray}{= \trace{ \mathcal{O}_\mathrm{EOM}^{(2)} [\partial_i U]_{-} [\partial_i \partial_j U]_{-} [\partial_j U]_{-} }} \,;
\end{align}
where $[A]_{-} \equiv \frac{1}{2} \left( A U^\dagger - U A^\dagger \right)$ for any $A$ transforming as $A \to L A R^\dagger$ under the group $SU(2)_L \times SU(2)_R$ and $\mathcal{O}_\mathrm{EOM}^{(2)} \equiv [\partial^2 U]_{-}$ defines the equation of motion at $\mathcal{O}(p^2)$, implying $\mathcal{O}_\mathrm{EOM}^{(2)}=0 + \mathcal{O}(p^4)$.

\bibliographystyle{style} 

\bibliography{refs} 

\end{document}